\documentclass[10pt]{article} 
\usepackage[preprint]{tmlr}

\usepackage{amsmath,amsfonts,bm}

\def\eqref#1{equation~\ref{#1}}

\def\1{\bm{1}}

\DeclareMathAlphabet{\mathsfit}{\encodingdefault}{\sfdefault}{m}{sl}
\SetMathAlphabet{\mathsfit}{bold}{\encodingdefault}{\sfdefault}{bx}{n}

\newcommand{\E}{\mathbb{E}}

\usepackage{hyperref}
\hypersetup{colorlinks=true,linkcolor=blue!50!black,citecolor=blue!50!black,urlcolor=blue!50!black}
\usepackage{url}
\usepackage{graphicx}
\usepackage{booktabs}
\usepackage{longtable}
\usepackage{amsthm}
\usepackage{amssymb}
\usepackage{xcolor}
\usepackage{enumitem}
\usepackage{placeins}
\setlist{itemsep=2pt,topsep=3pt}

\newtheorem{proposition}{Proposition}
\newtheorem{corollary}{Corollary}
\theoremstyle{definition}
\newtheorem{definition}{Definition}

\definecolor{boxgray}{RGB}{245,246,250}
\newcommand{\takeaway}[2]{\par\smallskip\noindent\colorbox{boxgray}{\begin{minipage}{0.97\linewidth}\smallskip\textbf{#1}\quad #2\smallskip\end{minipage}}\par\smallskip}
\newcommand{\Sb}{\bar S}

\title{The Limits of Query Monitoring against\\ Distributed Model Extraction}

\author{\name Snigdha Chandan Khilar \email snkhilar@gmail.com \\
      \addr Independent Researcher}

\newcommand{\coderepo}{\url{https://github.com/nssprogrammer/query-monitoring}}

\begin{document}

\maketitle

\begin{abstract}
Model providers now face distillation campaigns run through thousands of accounts, each of which looks like an ordinary user. When network and payment metadata cannot link those accounts, the only evidence left is what they ask. We study what a provider can and cannot learn from those queries alone, using one million real chat turns for development and 17 later weeks that no design choice touched for a single frozen evaluation. The answer is set less by anything specific to extraction than by benign coordination: a campaign is visible only if it is more distinctive than the most distinctive group of ordinary users in the same week. We make this precise, show that a small but positive difference between campaign and users buys a detection \emph{time} rather than detectability, and identify the campaign that is hardest to detect. Empirically, a calibrated monitor catches naive distributed extraction that per-account monitors miss (0.86--1.00 of weeks), and one variant---selecting accounts that move together after discounting how topical each one is---still catches campaigns padded with natural text (0.50--0.96 of weeks) where the other two spectral detectors catch none. Distilling each campaign into a small student shows that, in our testbed, campaigns that evaded every detector kept at most 42\% of the capability of a detectable one. The limits are equally concrete: a campaign that ramps up slowly poisons every detector's calibration, benign evaluation pipelines look exactly like distillation, we found no deployable detector at colluder shares of 2.5\% or below, and campaigns that relay real users' requests may be invisible to any query-only monitor. Every prediction we locked before each run, including the 74 that failed, is released with the code at \coderepo.
\end{abstract}

\section{Introduction}
\label{sec:intro}

A frontier model is expensive to train and cheap to query. That asymmetry makes \emph{distillation through the API} attractive: collect the model's answers to many well-chosen prompts and train a smaller model to imitate them \citep{hinton2015distilling,tramer2016stealing,orekondy2019knockoff}. Providers have long defended against this by flagging individual accounts whose queries look unusual \citep{juuti2019prada,zhang2021seat,sadeghzadeh2023hoda}. A capable attacker defeats that defence by spreading the campaign over many accounts, each of which then looks like an ordinary user. This is happening: in 2026 Anthropic reported distillation run through about 24{,}000 fraudulent accounts \citep{anthropic2026distillation} and campaigns that relayed real users' requests \citep{anthropic2026threat}; Google reported a campaign of more than 100{,}000 prompts \citep{google2026gtig}; and a joint U.S.\ government advisory described industrial-scale campaigns with automated failover to alternate access pathways \citep{cisa2026aa26251a}. Providers first try to \emph{link} such accounts through network, payment and device metadata, but a determined attacker can scrub those signals.

\paragraph{The question.} Suppose linkage has failed. The provider still sees every query from every account. \emph{Can it tell, from the queries alone, that a group of seemingly unrelated accounts is jointly extracting its model---and if so, when, at what false-alarm rate, and at what cost to the attacker?}

\paragraph{The short answer.} Query monitoring is worth running, but it is a filter, not a wall. What decides whether a campaign is visible is not a signature of extraction but \emph{benign coordination}: ordinary users also cluster on topics, follow trends, run evaluation scripts and paste templates. A campaign is visible only if it stands out more than the most distinctive group of ordinary users in the same week. We call this the \emph{benign-coordination barrier}; nearly every result below, positive or negative, follows from it.

\begin{figure}[t]
\centering
\includegraphics[width=\linewidth]{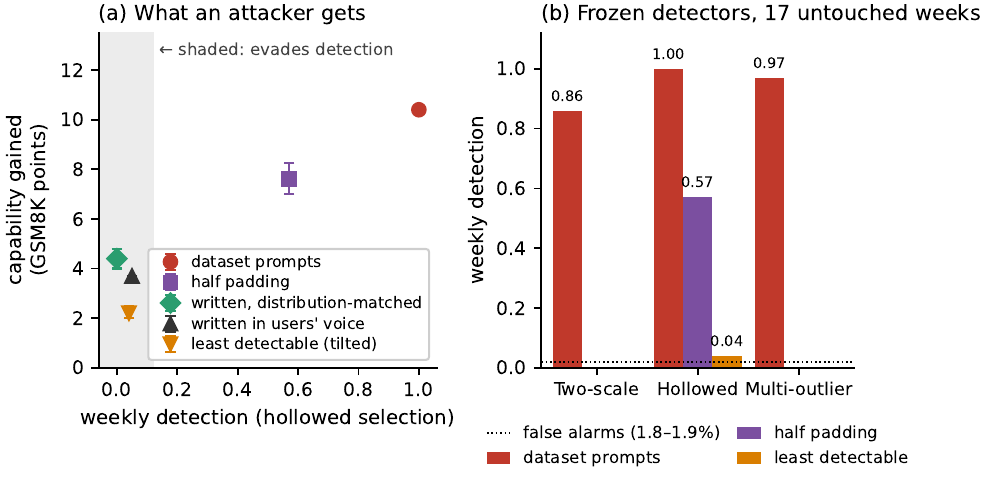}
\caption{The paper at a glance. (a) Capability a small student gains from 3{,}000 queries of each campaign type (GSM8K points; bars span two seeds) against how often hollowed selection, our most robust detector, catches the campaign each week (untouched weeks for the dataset, padded and least-detectable campaigns; development weeks for the two written ones). Every campaign in the shaded band evaded detection and kept at most 42\% of the dataset campaign's gain. (b) Three frozen detectors on 17 weeks that no design choice had touched; missing bars are zero. Only hollowed selection still catches campaigns padded with natural prompts; none catches the least-detectable campaign.}
\label{fig:glance}
\end{figure}

\paragraph{What this offers a provider.} We built and stress-tested a complete query-only monitor on real chat traffic \citep{zhao2024wildchat}, and distilled every campaign into a small student so that detection could be weighed against what the attacker gains (Figure~\ref{fig:glance}):
\begin{enumerate}[leftmargin=1.5em]
\item \textbf{Naive distributed extraction is catchable.} Dataset-like campaigns spread over many volume-matched accounts were caught in 86--100\% of weeks on data the monitor had never seen, at about 2\% false alarms; per-account monitors, tested on the development weeks, caught essentially none.
\item \textbf{The obvious evasion can be countered.} Padding every colluder account with natural prompts defeated our first detector. Judging accounts by whether they \emph{move together}, after discounting how topical each one is, still caught padded campaigns in 50--96\% of untouched weeks.
\item \textbf{Covertness was expensive in our testbed.} Every campaign that evaded every detector kept at most 42\% of what a detectable campaign transferred to a student that lacked the skill.
\item \textbf{The key quantity is measurable.} How coordinated a provider's benign traffic is decides what any monitor can see; it doubled between two periods of the same log, and we provide a script to measure it.
\end{enumerate}

\paragraph{What it cannot do.} Campaigns that relay real users' requests may have the same statistical law as real users, and then no query-only monitor can separate them. A campaign that ramps up slowly through the calibration weeks sharply cut the detection rate of every detector we built, two of them to near zero. At colluder shares of 2.5\% or below we found no detector that was both calibrated and powerful. And the monitor detects coordinated topic shift, not extraction: a customer benchmarking a model from ten accounts looks like a distiller.

\paragraph{How we kept ourselves honest.} Every experiment had predictions written into its notebook before it ran; of 184, 74 failed, and the full log is released with the code (\coderepo) together with the corrections we made along the way. Because later design choices were tuned on the same 24 development weeks, we froze the pipeline and evaluated it once on 17 later weeks from a different release of the log (Section~\ref{sec:holdout}); where the two disagree, the untouched weeks decide.

\paragraph{Contributions.} (i) A formal account of the benign-coordination barrier: law-matching campaigns are undetectable, a positive gap buys a detection \emph{time}, and the hardest campaign to detect is an exponential tilt of benign traffic whose capacity obeys a square-root law (Section~\ref{sec:theory}). (ii) A query-only monitor whose calibration survives weekly drift, with a hollowed variant that resists padding (Sections~\ref{sec:monitor} and~\ref{sec:results}). (iii) A capability--detectability frontier measured by distillation (Section~\ref{sec:frontier}). (iv) A frozen evaluation on untouched weeks with a second encoder (Section~\ref{sec:holdout}). (v) Code that reproduces every number, with each number mapped to the line that computes it and a complete record of predictions and corrections (Appendix~\ref{app:repro}).

\section{The setting}
\label{sec:setting}

A provider monitors $A$ accounts over a one-week \emph{window}. An attacker controls $K$ of them (the \emph{colluders}) and spreads $N$ queries across them; the rest are \emph{benign} users. The provider does not know $K$ or which accounts collude and has no linking metadata, but it has an embedding of every query and the history of past weeks. It wants a test that fires in weeks containing a campaign, fires in at most $\alpha=1\%$ of weeks without one, and names a short list of accounts for review. Each colluder sends as many queries as a typical benign account that week, so no single account is unusual; the signal, if any, lies in the joint structure of many accounts.

\emph{Deployed} detection uses only thresholds set in advance from past weeks. \emph{Oracle} detection sets the threshold on the null weeks being tested and measures only how well a statistic ranks campaigns; we label it every time. Each campaign window is scored with the identical window without the campaign, and a detection counts only if the detector fires with the campaign and not without it. (In the development-week experiments from hollowed selection onward, the paired null was another null window of the same week instead; with realised false alarms of about 1\% this changes paired detection by at most about that much, and the untouched-week evaluation uses identical-window pairing.)

\paragraph{Data.} We use English user turns from WildChat \citep{zhao2024wildchat}, treating each hashed IP as an account---a proxy that merges users behind shared addresses---in consumer chat rather than API traffic. \emph{Development weeks}: WildChat-1M, about one million turns from 98{,}331 hashed IPs over 56 weeks (April 2023--April 2024); every design choice was made here, and the last 24 weeks are test weeks. \emph{Untouched weeks}: WildChat-4.8M from 6 May to 1 September 2024 (17 weeks), with the 31 preceding weeks as calibration history only. A random 20\% of IPs supplies the attacker's pool of natural prompts. Monitored accounts have at least 10 queries in the week, and windows hold 64--364 accounts. The $K=50$ colluders replace benign accounts, and windows with fewer than 101 accounts are skipped when a campaign is injected, so colluders are 14--50\% of a window---far above the 0.05\% that 50 colluders would be among $10^5$ production accounts. Campaign prompts come from MetaMathQA \citep{yu2024metamath}, Magicoder \citep{wei2024magicoder}, Alpaca \citep{taori2023alpaca}, human-written GSM8K questions \citep{cobbe2021gsm8k}, the attacker's pool of real prompts, and LLM rewrites in the voice of real users.

\section{How a query-only monitor works}
\label{sec:monitor}

\begin{figure}[t]
\centering
\includegraphics[width=\linewidth]{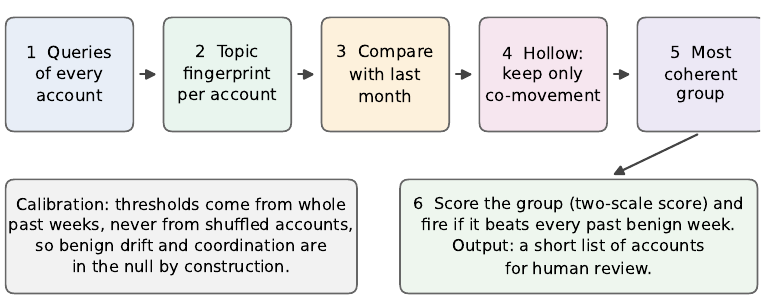}
\caption{The monitor. Each account's queries become a topic fingerprint, compared with last month's traffic; hollowing keeps only how accounts co-vary; the most coherent group is scored against every past benign week.}
\label{fig:monitor}
\end{figure}

Each week the monitor asks: \emph{is there a group of accounts whose queries are jointly unusual, compared with how groups of ordinary users looked in past weeks?} Figure~\ref{fig:monitor} shows the pipeline; Appendix~\ref{sec:monitor-full} gives details.

\paragraph{Fingerprints and whitening.} Each query is embedded (MiniLM; \citealp{wang2020minilm,reimers2019sbert}). An account's queries $x_1,\dots,x_{n_a}$ are summarised by random Fourier amplitudes $\rho_a(\omega_j)=\sum_i e^{-\mathrm{i}\,\omega_j\cdot\varphi(x_i)}$ \citep{rahimi2007random}: each random frequency probes one direction and scale of topic space, and an account that keeps returning to one region produces large, aligned amplitudes. Each account is then \emph{whitened} against the benign traffic of the four preceding weeks, $y_a=\Sigma^{-1/2}(\rho_a-n_a\mu)/\sqrt{n_a}$, so that an account sending ordinary queries has identity covariance however popular its topics are. The lagged reference absorbs slow drift.

\paragraph{Groups that move together.} Stacking the $y_a$ into a matrix $Y$ gives the cross-account matrix $C=YY^\top/r$. Independent accounts produce a random-matrix spectrum \citep{marchenko1967distribution}; a campaign whose accounts jointly concentrate adds a spike whose eigenvector points at the colluders \citep{baik2005phase}. The detector takes the fewest accounts carrying half of the top eigenvector's weight as its candidate group $G$ and scores it with the \emph{two-scale score} $T_{\rm 2s}(G)=\log\Sb_{h=4}(G)-\log\Sb_{h=0.5}(G)$, where $\Sb_h$ is the group's concentration at a coarse or fine bandwidth relative to independent queries. A group concentrated on a topic without the near-duplicate clumping of benign groups scores high. Because the score is a ratio within one group, weekly drift and population size largely cancel.

\paragraph{Hollowing.} $C$ has a weakness that turned out to matter: its diagonal holds each account's own concentration, so an individually topical benign user can dominate the top eigenvector, and padding a colluder with natural text shrinks its contribution. The \emph{centred, hollowed} matrix
\[
R_c=\Delta^{-1/2}\bigl(\tilde Y\tilde Y^\top-\Delta\bigr)\Delta^{-1/2},\qquad \Delta=\mathrm{diag}\bigl(\|\tilde y_a\|^2\bigr),
\]
where $\tilde Y$ is $Y$ with its mean row removed, deletes each account's self-term, normalises every account to unit length and removes the week's common drift \citep{cai2021subspace,zhang2022heteroskedastic}. Its top eigenvector picks accounts whose deviations \emph{point the same way}, however large or small. We call selection on $R_c$ followed by the two-scale score \emph{hollowed selection}. \emph{Multi-outlier selection} scores a group from every eigenvector of $R_c$ above the random-matrix bulk (up to 20) and keeps the largest score, $T_{\max}$.

\paragraph{Calibration.} We never build the null by shuffling accounts or queries, which destroys exactly the benign coordination the test must tolerate. Each threshold is the $(1-\alpha)$ quantile of the same statistic over \emph{whole} past weeks---the eight preceding weeks (\emph{rolling}) or all prior weeks (\emph{conformal})---each whitened against its own lagged reference. The spectral step costs $O(r^3)$ for $r$ whitened components, independent of the number of accounts; a weekly window took about 0.2\,s on one T4 GPU.

\paragraph{Baselines.} A whole-window shift test (the \emph{global stream}), the MMD \citep{gretton2012kernel} traffic-window test of \citet{liu2026embarrassingly}, per-account and Higher Criticism tests \citep{donoho2004hc}, Fraudar \citep{hooi2016fraudar}, a volume test, classifiers and topic statistics on the selected group, and a scan of topic-cell volumes, all with the same references and calibration.

\section{Why detection is hard: the benign-coordination barrier}
\label{sec:theory}

A monitor that looks for unusually coordinated groups will always find \emph{some} group, because ordinary users coordinate: a homework deadline sends hundreds of students to the same problems; a company's scripts send templated prompts from several accounts. A campaign is caught only if it is more distinctive than the most distinctive of these benign groups. The four results below make this precise. Each is a standard argument applied to this problem; their value is that they tell a provider what to measure. Proofs are in Appendix~\ref{app:proofs} and further discussion in Appendix~\ref{app:theory}.

\begin{definition}[Benign-coordination barrier]\label{def:barrier}
For a detector that scores groups of accounts by $\psi$ and fires when a week's highest score exceeds a threshold, the barrier is the $(1-\alpha)$ quantile of the most distinctive \emph{benign} group's score in a week of $A$ accounts, $B_\alpha(A)=\inf\{\tau:\Pr_0(\max_{G\ \text{benign}}\psi(G)\le\tau)\ge1-\alpha\}$. A campaign with score $\psi_c$ is detected exactly when $\psi_c>B_\alpha(A)$.
\end{definition}

For a max-type detector this nearly restates what a threshold is; its use is that $B_\alpha(A)$ is a property of benign traffic alone, measurable without any campaign.

\begin{proposition}[Law matching]\label{prop:law}
If a benign group of $K$ accounts appears in a week with probability $\pi>0$, and when it appears its accounts' queries have the same joint law as the colluders' and are independent of the rest of the week, then every test with false-alarm rate at most $\alpha$ has power at most $\alpha/\pi$.
\end{proposition}

If real users sometimes behave exactly like the campaign, catching the campaign means flagging them too. This is the indistinguishability argument of secure steganography \citep{hopper2002provably}; its message is a design principle: \emph{a lasting defence needs a feature in which campaigns and benign groups differ in law}. It is also why relays are the hard case (Section~\ref{sec:world}).

\begin{proposition}[More benign groups raise the bar]\label{prop:max}
If benign group scores are independent with survival function $\bar F$ and their number per week is Poisson with mean $\nu$, a level-$\alpha$ threshold satisfies $\bar F(\tau)\approx\alpha/\nu$: a campaign must out-rank all but a fraction $\alpha/\nu$ of benign groups.
\end{proposition}

Going from no tight benign cluster per week to one raises the bar abruptly, which is the collapse we see in a controlled study (Appendix~\ref{app:controlled}). Our real data refute the independent-groups model for the two-scale score, whose threshold stayed flat as populations grew, so we use this proposition only qualitatively.

\begin{proposition}[A positive gap buys time]\label{prop:seq}
If weekly scores are independent with law $P_0$ without a campaign and $P_1$ with a persistent one, the best sequential test with false-alarm rate $\alpha$ needs on average about $\ln(1/\alpha)/D(P_1\|P_0)$ weeks \citep{wald1945sequential,wald1948optimum}.
\end{proposition}

Mimicry that shrinks the gap $D$ therefore multiplies the time a defender must watch rather than making the campaign invisible. A conformal test martingale makes this deployable while keeping the probability of \emph{any} false alarm, ever, below $\alpha$ \citep{vovk2021testing,vovk2021retrain,ramdas2023gametheoretic}.

\begin{proposition}[The hardest campaign to detect]\label{prop:lfc}
Let $S$ be a region of useful queries with benign share $p$. Among query laws with at least a fraction $c>p$ in $S$, the closest to benign traffic in relative entropy is the exponential tilt $Q^*=c\,P_0(\cdot\mid S)+(1-c)\,P_0(\cdot\mid S^c)$, at $d(c\|p)=c\ln\frac{c}{p}+(1-c)\ln\frac{1-c}{1-p}$ per query \citep{csiszar1975idivergence}. No query-based test that knows which accounts to watch detects $N_w$ such queries per week in fewer than about $\ln(1/\alpha)/(N_w d(c\|p))$ weeks on average.
\end{proposition}

The least-detectable campaign is not ``dataset prompts that look natural'' but \emph{real benign prompts, re-weighted toward the useful region}. Expanding $d(c\|p)$ gives a square-root law \citep{ker2007capacity,bash2013limits}: over $n$ queries an attacker can add only about $\sqrt{2p(1-p)\,n\ln(1/\alpha)}$ useful queries beyond the $pn$ that benign users send anyway---about 51 over twelve weeks of 1{,}640 queries with the measured $p=1.48\%$. If the monitor does not know which accounts to watch, the attacker can instead hide among \emph{types} of users.

\begin{corollary}[Account types]\label{cor:types}
If a share $q$ of benign accounts are on-task users and each colluder poses as one, a monitor that sees every account's type but not which accounts collude needs more than about $K^*\approx\sqrt{2Aq\ln(1/\alpha)/(w(1-q))}$ posing accounts to detect the campaign over $w$ weeks.
\end{corollary}

Capacity is then counted in accounts and grows like $\sqrt A$; in WildChat on-task users are rare ($q=0.25\%$), so even at $A=10^5$ the bound allows only about 14 posing accounts over twelve weeks.

\section{What the monitor sees on development weeks}
\label{sec:results}

This section summarises the development weeks; Appendix~\ref{sec:results-full} gives the full results and Section~\ref{sec:holdout} re-tests the final pipeline on untouched weeks. Colluders are $K=50$ volume-matched accounts sending about 1{,}640 queries a week. Weeks, not windows, are the independent unit: zero false alarms in 960 windows from 24 weeks bounds the rate at about 12\%, not 0.3\%. Run-to-run variation (0.79--0.93 for one cell across four runs) exceeds single-run intervals, so we compare detectors within a run.

\paragraph{Calibration breaks first.} Thresholds set once on early weeks fired on 12\% of later null weeks against a nominal 1\%, and building the null by resampling individual accounts fails for the same reason: it erases benign coordination. Resampling \emph{whole} past weeks against a lagged reference works. Over whole weekly populations only the two-scale test stayed calibrated under rolling calibration (0.7\% false alarms), against 2.9--14.2\% for every other detector (Appendix Figure~\ref{fig:calib-full}).

\begin{table}[t]
\caption{Weekly paired detection at deployed thresholds (nominal 1\%), whole weekly populations, $K=50$ volume-matched colluders, 24 development weeks. The last row is the realised false-alarm rate of each detector.}
\label{tab:deploy}
\begin{center}
\footnotesize\setlength{\tabcolsep}{4pt}
\begin{tabular}{lccccc}
\toprule
Campaign & Two-scale & Global stream & MMD & Spectral & Per-account / HC / Fraudar \\
\midrule
MetaMathQA (generated math) & \textbf{0.93} & 0.71 & 0.60 & 0.19 & $\le$0.01 \\
Magicoder (generated code) & \textbf{0.62} & 0.46 & 0.32 & 0.00 & 0.00 \\
Alpaca (generated, broad) & 0.01 & 0.00 & 0.01 & 0.00 & 0.00 \\
Narrow cluster of real prompts & 0.12 & 0.04 & \textbf{0.27} & 0.00 & 0.00 \\
Random real prompts & 0.01 & 0.00 & 0.01 & 0.00 & 0.00 \\
\midrule
Realised false alarms & 0.7\% & 4.1\% & 11.6\% & 7.8\% & 5.3 / 7.5 / 2.9\% \\
\bottomrule
\end{tabular}
\end{center}
\end{table}

\paragraph{Coordinated, dataset-like campaigns are catchable.} Topically focused generated datasets were caught in most weeks, while no per-account monitor caught anything (Table~\ref{tab:deploy}). Broad datasets and real user prompts were not---exactly what the barrier predicts. When the two-scale test fired, the flagged group was almost entirely colluders (precision 0.94--1.00) but held only 12--31\% of them: a pointer for review, not a full list. Its edge is calibration rather than ranking: at oracle-matched 1\% false alarms it caught 0.80 of MetaMathQA campaigns against 0.70 for the global stream and 0.58 for MMD, but the global stream ranks slightly better overall (AUC 0.97 against 0.88).

\paragraph{It detects topic shift, not extraction.} We designed the two-scale test on the idea that it would detect deduplicated generated data. Three findings rejected that idea. Coordinated \emph{human-written} GSM8K questions were caught more often (0.94) than generated MetaMathQA (0.79), while simpler source-shift statistics on the same group---a dataset-versus-chat classifier, topic concentration---caught nothing at 1\%. And benign groups sweeping the GSM8K test set from 3 or 10 accounts triggered the test in 79--80\% of weeks, as often as distillation.

\paragraph{Padding, and a fix that looks at directions.} Mixing natural prompts into every colluder account cut two-scale detection from 0.85 to 0.21--0.24. Four repairs that pool topic cells across accounts ranked padded campaigns better but never reached a usable false-alarm rate. What worked was asking \emph{why} padding works: it shrinks the \emph{size} of each colluder's deviation but not its \emph{direction}. With conformal thresholds, hollowed selection caught padded campaigns 0.46--0.48 of the time, against 0.00--0.02 for selection on $C$, with no false alarm in 960 windows. Once topicality and drift are removed, benign accounts in these weeks were barely more coordinated than chance: the top eigenvalue of $R_c$ was 1.14 times that of shuffled weeks.

\paragraph{Scale: selection fails before the barrier does.} At colluder shares of 0.4--2.5\% (populations of 1{,}000 and 2{,}500 real account-weeks) no detector was both calibrated and powerful. In populations pooled from account-weeks across the year (Appendix Figure~\ref{fig:scale-full}), two-scale power fell from 1.00 to 0.00 as the colluder share went from 20\% to 0.5\%, yet the \emph{true} colluder set still scored above the threshold at every size: the detector stopped finding the right accounts. Multi-outlier selection closed that gap with oracle thresholds (93--100\% of unpadded campaigns up to $10^4$ accounts), but its best deployed calibration reached 5\% false alarms, not 1\%, and pooling probably understates same-week benign coordination. \textbf{For deployment at realistic colluder shares, our result is negative.} Query monitoring is much stronger as a complement to metadata linkage: given 30\% of a campaign's colluders, expanding along $R_c$ recovered 96--100\% of all colluders (oracle thresholds, known campaign size).

\paragraph{Attacks on the monitor.} Ordering prompts greedily against the two-scale statistic cut its detection from 0.80 to 0.31 without any padding. Giving the most distinctive queries to five sacrificial accounts left selection on $C$ flagging those five and none of the other 45 colluders. And ramping a campaign up through the eight calibration weeks poisoned the reference and the thresholds, cutting detection from 0.95 to 0.12---the boiling-frog attack on anomaly detectors \citep{rubinstein2009antidote}.

\section{What the attacker gets: the capability--detectability frontier}
\label{sec:frontier}

An attacker who evades the monitor but learns nothing has not won. To weigh detection against gain, we turned every campaign into exactly the (prompt, answer) pairs it would collect---dataset answers for dataset prompts, the recorded ChatGPT replies for WildChat prompts---fine-tuned a small student on them, and measured GSM8K accuracy \citep{cobbe2021gsm8k}. Appendix~\ref{sec:frontier-full} gives the full protocol and Appendix Figure~\ref{fig:stealth-full} the gain curves.

\paragraph{Two regimes.} Students that already have the skill learn only the answer format. Qwen2.5-0.5B \citep{qwen2024qwen25} gained 9.8 points from 100 math prompts and no more from 900, and padded or natural-looking prompts did as well; Qwen2.5-1.5B on MATH-500 was flat from 1{,}000 to 4{,}000 queries. In this \emph{elicitation} regime, consistent with \citet{gudibande2024false}, stealth is free. A price of stealth can appear only when the student \emph{lacks} the skill. SmolLM2-360M \citep{allal2025smollm2} starts at 3.5\% and gained 7.5, 10.4 and 15.9 points from 1{,}000, 3{,}000 and 10{,}000 MetaMathQA queries---the \emph{extraction} regime. Padding half of every campaign with natural prompts cost 3.6, 2.8 and 3.8 points (48\%, 27\% and 24\% of the gain): the natural prompts taught nothing, so the cost of mimicry is the queries spent on it.

\paragraph{The covert end.} Proposition~\ref{prop:lfc}'s least-detectable campaign---half real WildChat prompts in the math region, half the rest---evaded every detector and gained 2.2 points at 3{,}000 queries, against 10.4: covertness cost 79\% of the gain. To test whether WildChat's math is simply the wrong math, we let the attacker \emph{write} useful prompts in users' voices. An instruction-tuned model rewrote GSM8K-style problems in the style of real WildChat messages, but a classifier still told the rewrites from real prompts with AUC 0.97--0.98, and only 7.9\% passed as user-like. Resampling all rewrites to match real on-task prompts as closely as the classifier could measure gave a campaign that evaded every detector and gained 4.4 points---42\% of the dataset gain, the best any covert campaign achieved---from an effective sample of only 151 distinct rewrites.

\takeaway{The frontier, in one sentence.}{In our testbed, every campaign that evaded every detector also gave up 58\% or more of what a detectable campaign transferred, and each was limited by its supply of prompts that are both useful and indistinguishable from users'.}

This is narrower than it sounds: one 360M student, one benchmark, answers from datasets rather than a live teacher, and two seeds; the covert gains of 1.1--4.4 points are differences of 4--18 of 400 problems. And attackers with a user base do not face the supply limit (Section~\ref{sec:world}).

\paragraph{Capability held at first detection.} What an attacker optimises is the capability it holds when first caught. Table~\ref{tab:timefrontier} combines the gain curves with detection times at 1{,}640 queries a week under two rules applied identically to every detector: the \emph{first weekly alarm} at a 1\% threshold, which is fast but accumulates false alarms (null runs alarmed within 16 weeks a quarter of the time on untouched weeks), and the \emph{anytime-valid} conformal test martingale, which holds the probability of any false alarm, ever, at 1\%.

\begin{table}[t]
\caption{Capability held at first detection: GSM8K gain (points) of the extraction-regime student at the number of queries sent by the median week of first detection, at 1{,}640 queries a week. Each cell: development weeks / untouched weeks. ``none'': no alarm within 16 weeks.}
\label{tab:timefrontier}
\begin{center}
\footnotesize\setlength{\tabcolsep}{3.5pt}
\begin{tabular}{llccc}
\toprule
Detector & Campaign & Weekly detection & First weekly alarm & Anytime-valid \\
\midrule
Two-scale & Dataset & 0.79--0.85 / 0.86 & week 1 (8.8) / week 1 (8.8) & week 3 (12.7) / week 3 (12.7) \\
 & Half padding & 0.21 / 0.00 & week 3 (9.5) / none & week 15 (15.5) / none \\
Hollowed & Dataset & 1.00 / 1.00 & week 1 (8.8) / week 1 (8.8) & not run \\
 & Half padding & 0.44--0.48 / 0.50--0.57 & week 2 (7.9) / week 1 (5.6) & not run \\
Multi-outlier & Dataset & 1.00 / 0.97 & week 1 (8.8) / week 1 (8.8) & not run / week 3 (12.7) \\
 & Half padding & 0.36--0.42 / 0.00 & week 2 (7.9) / none & not run / 25\% of runs \\
Any & Least detectable & $\approx$0.01 / $\le$0.04 & \multicolumn{2}{c}{indistinguishable from false alarms} \\
\bottomrule
\end{tabular}
\end{center}
\end{table}

Against the two-scale and multi-outlier tests, padding on the untouched weeks bought evasion for at least a quarter of a year in most runs at a cost of a quarter to a half of the gain. Against hollowed selection it bought nothing: padded campaigns were caught in the first week. The time to an anytime-valid alarm for hollowed selection is the most important number we did not measure.

\section{The test that matters: one frozen evaluation on untouched weeks}
\label{sec:holdout}

Every choice behind hollowed and multi-outlier selection---hollowing, centring, conformal thresholds, the bulk-edge factor of 1.05, the minimum group spread of 5 accounts, the cap of 20 eigenvectors---was made after seeing results on the same 24 development weeks. Such results are optimistic by an amount those weeks cannot estimate. We therefore froze the pipeline and ran it once, with 13 predictions locked beforehand, on 17 later weeks of WildChat-4.8M, fitting the PCA basis, topics, task classifier and every threshold on the 31 preceding history weeks only. We repeated everything with a second encoder, E5-small-v2 (initialised from MiniLM, so not independent), and in whole weekly populations. Appendix~\ref{sec:holdout-full} gives details.

\begin{table}[t]
\caption{The frozen pipeline on 17 untouched weeks: weekly paired detection at deployed thresholds, nominal 1\% (conformal over all prior weeks; eight preceding weeks for the volume scan). Development-week values for comparison: dataset 0.79--0.93 / 1.00 / 1.00; half padding 0.21 / 0.44--0.48 / 0.36--0.42.}
\label{tab:holdout}
\begin{center}
\footnotesize\setlength{\tabcolsep}{3.5pt}
\begin{tabular}{llcccc}
\toprule
Setting & & Two-scale & Hollowed & Multi-outlier & Volume scan \\
\midrule
MiniLM, 64--252 accounts & False alarms & 1.8\% & 1.9\% & 1.9\% & 1.5\% \\
 & Dataset (MetaMathQA) & 0.86 & 1.00 & 0.97 & 0.61 \\
 & Half padding & 0.00 & \textbf{0.57} & 0.00 & 0.37 \\
 & Least detectable (tilted) & 0.00 & 0.04 & 0.00 & 0.10 \\
 & Random real prompts & 0.00 & 0.00 & 0.00 & 0.00 \\
 & GSM8K sweep, 10 / 3 accounts & 0.87 / 0.88 & 1.00 / 0.34 & 0.97 / 0.93 & 0.46 / 0.46 \\
\midrule
E5-small-v2 & False alarms & 1.3\% & 0.3\% & 0.0\% & 6.2\% \\
 & Dataset & 0.86 & 1.00 & 1.00 & 0.48 \\
 & Half padding & 0.00 & \textbf{0.89} & 0.00 & 0.09 \\
 & Least detectable & 0.00 & 0.03 & 0.00 & 0.02 \\
\midrule
MiniLM, whole weeks & False alarms & 1.0\% & 0.5\% & 2.5\% & 11.3\% \\
(260--469 accounts) & Dataset & 0.90 & 1.00 & 1.00 & 0.84 \\
 & Half padding & 0.00 & \textbf{0.96} & 0.00 & 0.51 \\
\bottomrule
\end{tabular}
\end{center}
\end{table}

\paragraph{What held.} Calibration held approximately: 1.8--1.9\% false alarms against a nominal 1\% with MiniLM and 0--1.3\% with E5, so a provider should budget for about twice the nominal rate. Dataset campaigns were caught 0.86--1.00 of the time with both encoders, and the least-detectable campaign and random real prompts stayed at or below 0.04 for the three spectral detectors (Table~\ref{tab:holdout}). Most importantly, \textbf{hollowed selection still saw through padding}: 0.50--0.57 across three campaign seeds, 0.89 with E5, 0.96 in whole weekly populations. It was also unaffected by sacrificial scheduling, still flagging 49--59\% of the non-sacrificial colluders: hollowing normalises each account, so concentrating distinctive queries on a few accounts does not move the direction the colluders share.

\paragraph{What did not.} Under padding, the two-scale test fell from 0.21 to 0.00 and multi-outlier selection from 0.36--0.42 to 0.00. Multi-outlier selection pays inside its threshold for taking a maximum over up to 20 groups, and here the colluders' group no longer out-scored the most distinctive benign groups, even though the hollowed top eigenvector still found it; it should run beside, not instead of, hollowed selection. Benign coordination was higher than before: the real-to-shuffled eigenvalue ratio of $R_c$ was 2.22 (MiniLM) and 3.06 (E5), against 1.14. The calibration ramp was far worse than predicted, cutting multi-outlier detection from 0.89 to 0.00, the conformal two-scale test from 0.80 to 0.02 and hollowed selection from 0.98 to 0.32: a 1\% conformal threshold is the 99th percentile of past null scores, so once contaminated windows exceed 1\% of the history they set it.

\paragraph{Covert is not invisible.} Hollowed selection \emph{ranked} the least-detectable campaign well above benign windows (AUC 0.97 MiniLM, 0.92 E5) without detecting it at deployed thresholds. Random real prompts also ranked high (AUC 0.88), so part of this signal comes from how we build colluder accounts from pooled traffic. Whether a sequential test on the hollowed score catches covert campaigns, and whether real same-week populations of $10^4$ accounts are harder than our pooled ones (this period's weeks held at most 469 accounts), remain open.

\takeaway{What the untouched weeks changed.}{Hollowed selection is the most robust deployed detector we built: near-nominal false alarms, unaffected by sacrificial scheduling, and the only one that still catches padded campaigns at near-nominal false alarms in every setting. Benign coordination varies over time and must be measured. Calibration poisoning by a slow ramp is the most serious threat we found to every detector.}

\section{Where real campaigns sit, and what passive monitoring cannot do}
\label{sec:world}

\paragraph{Documented campaigns.} The campaigns reported in 2026 occupy two corners, neither of them the covert, self-written corner we studied most. At one are high-volume campaigns: the largest Anthropic has reported ran more than 151 million exchanges in three months through more than 3{,}500 accounts, several hundred exchanges per account per day, and its accounts were linked because they all used one fixed prompt \citep{anthropic2026threat}. Such campaigns are visible to per-account volume monitors and through their shared template, a kind of coordination our detectors caught almost every week. At the other corner are \emph{relays}: one forwarded almost 300{,}000 customer requests in ten days through 5{,}380 accounts, about six per account per day \citep{anthropic2026threat}.

\paragraph{Why relays are the hard case.} Our covert campaigns were limited by supply: an attacker without users could not obtain or write enough prompts that are both useful and user-like. A relay has an unlimited supply. If a relaying account's traffic has the same joint law as the users it relays, Proposition~\ref{prop:law} leaves no query-only monitor able to separate it. That ``if'' is an assumption: a relay that interleaves many users' sessions in one account, or forwards only high-value requests, may differ in session coherence or topicality. We did not simulate relays; it is the most important open experiment.

\paragraph{The provider's own randomness.} Every threshold here is learned from history, which drifts and can be poisoned. A provider has one resource no attacker controls: if it throttles a flagged group in randomly chosen time blocks, it can test whether the work fails over to other accounts, with a null that follows from the randomisation alone, whatever benign traffic does (Fisher's randomisation test in a switchback design; \citealp{fisher1935design,bojinov2023design}; Proposition~\ref{prop:rand} in Appendix~\ref{sec:world-full}). In simulation on WildChat populations the null held on real, drifting traffic, and campaigns that reassigned throttled work were caught in every run. But a benign team with legitimate failover fired the test 82--98\% of the time, throttling harms legitimate accounts in a flagged group, and twelve weekly blocks take a quarter of a year (Appendix~\ref{app:randomized}).

\section{Guidance for providers}
\label{sec:guidance}

Each item is backed by a measurement in this paper; Appendix~\ref{sec:guidance-full} gives the caveats and Appendix Table~\ref{tab:regime-full} what worked where.
\begin{enumerate}[leftmargin=1.5em,itemsep=1pt,topsep=2pt]
\item \textbf{Measure your benign coordination first:} per week, the real-to-shuffled ratio of the top eigenvalue of $R_c$, the number of outlier eigenvectors, and the share of accounts of each valuable type. These decide what any query-only monitor can see, and the eigenvalue ratio doubled between two periods of the same public log. A script that computes them is in the code release (\coderepo).
\item \textbf{Calibrate on whole past weeks against a lagged reference,} never by shuffling accounts or queries; budget for about twice the nominal false-alarm rate.
\item \textbf{Protect the reference and calibration history.} A slow ramp sharply cut the detection of every detector; keep frozen copies beside the rolling ones and audit threshold drift.
\item \textbf{Monitor whole populations and select on the hollowed matrix;} run multi-outlier selection beside it, not instead of it.
\item \textbf{Accumulate evidence over weeks} with a conformal test martingale: while the gap to benign traffic is positive, mimicry slows detection rather than preventing it.
\item \textbf{Combine with linkage in both directions,} and treat an alarm as a high-precision pointer for human review: the monitor detects coordinated topic shift, so evaluation pipelines will appear in the alarms.
\end{enumerate}

\section{Related work}
\label{sec:related}

Appendix~\ref{sec:related-full} gives a fuller account. \textbf{Extraction and its defences.} Extraction through prediction APIs \citep{tramer2016stealing,orekondy2019knockoff,pal2020activethief} and distillation from generated instruction data \citep{hinton2015distilling,taori2023alpaca,yu2024metamath} establish the threat. Most deployed defences analyse one account at a time \citep{juuti2019prada,zhang2021seat,sadeghzadeh2023hoda}; stateful detection \citep{chen2020stateful,li2022blacklight} discusses attackers who split work across accounts, and FDINet tests pairs of clients for collusion \citep{yao2023fdinet}. Closest to us, \citet{liu2026embarrassingly} apply a benign-calibrated MMD test to traffic windows of embeddings, including windows that mix campaign queries with cover queries; it is our strongest baseline, and we extend that line to whole real populations, colluder shares down to 0.5\% and a measured price of evasion. Other defences avoid detection altogether, through honeypots \citep{dai2026knowledgetrap,wang2025honeypotnet} or seeded generation that degrades multi-account distillation \citep{diao2026lads}. Measured prices of evasion appear in \citet{liu2026embarrassingly} and \citet{zhang2021seat}; we add the price in an extraction regime on real traffic and its time-integrated form.

\textbf{Coordination, mimicry and poisoning.} Lockstep-group detection in fraud and bot detection uses singular vectors, dense blocks and timing \citep{prakash2010eigenspokes,beutel2013copycatch,hooi2016fraudar,cao2014synchrotrap}, and benign coordination is a known confounder \citep{vargas2020detection}. Mimicry attacks on intrusion detectors \citep{wagner2002mimicry} and boiling-frog poisoning \citep{rubinstein2009antidote} are the ancestors of our padding and ramp attacks. \textbf{Statistics.} Our spectral analysis rests on the BBP transition \citep{baik2005phase,benaych2011eigenvalues}; hollowing belongs to the repairs of eigenvector localisation on heavy vertices \citep{alt2023localized,le2017concentration}; the selection gap echoes statistical--computational gaps in sparse PCA \citep{berthet2013optimal}; Proposition~\ref{prop:lfc} is the I-projection \citep{csiszar1975idivergence} read as an attack, with the square-root law of covert communication \citep{ker2007capacity,bash2013limits}; and our sequential tools are conformal test martingales \citep{vovk2021testing}.

\section{Limitations and open tests}
\label{sec:limitations}

Appendix~\ref{sec:limitations-full} expands each point. \textbf{Data:} WildChat is consumer chat, not API traffic; hashed IPs only approximate accounts; activity thresholds exclude low-volume accounts, where a patient attacker would hide; and colluders are 14--50\% of a window, far above production shares. \textbf{Representation:} attacker and defender share one family of sentence embeddings (E5-small-v2 is initialised from MiniLM); lexical, style, timing and victim-model features are untested. \textbf{Statistics:} conclusions rest on 17--24 test weeks; development results reflect adaptive reuse of the same weeks, which one frozen evaluation only partly corrects; predictions were locked in notebooks but carry no external timestamp. \textbf{Campaign construction:} colluder accounts were assembled from draws across a pool rather than from whole real user-weeks, which confounds the natural and covert results in an unknown direction. \textbf{Capability:} the frontier rests on one 360M student, one benchmark, answers from datasets rather than a live teacher, and two seeds; the imitating writer was a 1.5B model.

As future work, the experiments that would most change our conclusions are an anytime-valid sequential test on the hollowed score, colluders built from whole real user-weeks, a relay simulation, a defence against calibration poisoning, distillation from a live teacher with more seeds, students and tasks, and, above all, the benign-coordination census on a real provider's logs.

\section{Conclusion}
\label{sec:conclusion}

When accounts cannot be linked, what a provider can learn from queries alone is decided by benign coordination: a campaign is visible only if it stands out more than the most distinctive group of ordinary users. That one idea explains why naive distributed extraction is catchable, why padding blinds detectors that weigh accounts by size but not one that weighs them by direction, why selection fails before the barrier as populations grow, and why relays are the hard case.

For a provider, the result is a filter worth running and a clear list of what it will not do. A calibrated, hollowed monitor catches dataset-like campaigns spread across many accounts, keeps catching them when they are padded, and points reviewers at a short list of accounts that are mostly colluders; in our testbed, the campaigns it could not see had given up most of their value. But it detects topic shift, not extraction; a slow ramp badly degrades it; we found no deployable configuration at production colluder shares; and relays may be beyond any passive monitor. The decisive quantity---how coordinated a provider's own benign traffic is---can only be measured on real logs, and we hope the census script, the calibration protocol and the record of what failed make that measurement easy to repeat.

\subsubsection*{Broader Impact Statement}
This work aims to help model providers defend against unauthorised distillation, and it also describes ways to defeat the defences it studies: padding colluder accounts with natural text, ramping a campaign through the calibration weeks, choosing which accounts the detector flags, and re-weighting real user prompts toward a useful region. We report them because a defender cannot evaluate a monitor without them, and because each follows directly from the monitor's design or from long-known ideas (mimicry and boiling-frog poisoning); documented attackers already go further by relaying real users' requests at volumes far beyond ours. We release all of our code openly (\coderepo), including the detectors, the calibration protocol, the census script and the campaign constructions, so that defences can be evaluated against them.

\emph{Data.} WildChat-1M and WildChat-4.8M are public under the ODC-BY license, with accounts released as hashed IP addresses. We did not attempt to re-identify any user, inspected conversations only through automated processing, and reproduce no user prompt; every result is an aggregate over windows of tens to thousands of accounts.

\emph{Named organisations.} We name companies only as providers' and governments' public reports name them, describe their conduct as reported allegations, and make no independent claim about them.

\emph{Harms of deployment.} Population-level monitoring pools all users' queries into one statistic and can open side channels \citep{debenedetti2024privacy}; the monitor flags legitimate evaluation pipelines as readily as distillation; and randomised throttles degrade service for any legitimate account in a flagged group. Enforcement should follow human review, under the governance any population-level telemetry needs.

\bibliography{refs}
\bibliographystyle{tmlr}

\appendix
\section{Proofs}
\label{app:proofs}

\paragraph{Proposition~\ref{prop:law}.} By the assumptions (the matching group has the colluders' joint law and is independent of the rest of the week), an attacked week has the law of a benign week conditioned on the matching group $G$ appearing. Hence for any test $\phi$, $\Pr_{\rm attack}(\phi=1)=\Pr_0(\phi=1\mid G)\le \Pr_0(\phi=1)/\pi\le\alpha/\pi$.

\paragraph{Proposition~\ref{prop:max}.} With a Poisson($\nu$) number of i.i.d.\ benign group scores, the benign maximum is at most $\tau$ with probability $\E[(1-\bar F(\tau))^n]=\exp(-\nu\bar F(\tau))$; setting this to $1-\alpha$ gives $\bar F(\tau)=-\ln(1-\alpha)/\nu$. The week fires with the campaign but not without it exactly when $\psi_c>\tau$ and the benign maximum is at most $\tau$, so the paired detection probability is $(1-\alpha)\,\mathbf 1\{\bar F(\psi_c)<-\ln(1-\alpha)/\nu\}$. If benign group scores above a level $u$ follow a generalised Pareto tail with shape $\xi$, scale $\sigma$ and exceedance probability $p_u$, and $\nu=\kappa A$, substituting $\bar F(\tau)=p_u(1+\xi(\tau-u)/\sigma)^{-1/\xi}$ gives $B_\alpha(A)=u+\frac{\sigma}{\xi}\bigl[(p_u\kappa A/(-\ln(1-\alpha)))^{\xi}-1\bigr]$ \citep{pickands1975statistical,coles2001introduction}: logarithmic growth in $A$ for light tails and polynomial growth for heavy ones. Our data did not support this model for the two-scale score (its threshold stayed flat), so we use it only qualitatively.

\paragraph{Proposition~\ref{prop:seq}.} This is Wald's approximation to the expected sample size of the sequential probability ratio test, $\E_1[N]\approx[(1-\beta)\ln\frac{1-\beta}{\alpha}+\beta\ln\frac{\beta}{1-\alpha}]/D(P_1\|P_0)$, which ignores overshoot \citep{wald1945sequential}; optimality among tests with the same error rates is the Wald--Wolfowitz theorem \citep{wald1948optimum}. If the benign law may be any member of a family $\mathcal P_0$, the denominator becomes $\inf_{P\in\mathcal P_0}D(P_1\|P)$. A deployable version replaces the likelihood ratio by the running product of calibrated weekly e-values; with $n$ calibration windows the smallest conformal $p$-value is $1/(n+1)$, so one week contributes at most $\ln e(1/(n+1))$ to the log-martingale (about 2.24 nats with $n=320$ and the mixture calibrator $e(p)=(1-p+p\ln p)/(p\ln^2 p)$), which is why even strong campaigns take about three weeks at $\alpha=1\%$.

\paragraph{Proposition~\ref{prop:lfc}.} Let $u=\mathbf 1_S$, $\Lambda(\theta)=\ln\E_0[e^{\theta u}]$ and $Q_\theta$ the exponential tilt with parameter $\theta$. For any $Q$ with $\E_Q[u]\ge c$ and any $\theta\ge0$, $D(Q\|P_0)=D(Q\|Q_\theta)+\theta\E_Q[u]-\Lambda(\theta)\ge\theta c-\Lambda(\theta)$. Since $\Lambda'$ increases continuously from $p$ to 1, some $\theta^*$ has $\Lambda'(\theta^*)=c$, and the bound is attained at $Q_{\theta^*}$ only \citep{csiszar1975idivergence}. For an indicator, $Q_{\theta^*}=c\,P_0(\cdot\mid S)+(1-c)P_0(\cdot\mid S^c)$ and $\Lambda^*(c)=d(c\|p)$. An attacked week and the benign week in which the same accounts send draws from $P_0$ differ only in the law of $N_w$ independent queries, so their relative entropy is $N_w d(c\|p)$, and no feature of the queries carries more (data processing); Wald's lower bound then applies. For $c$ near $p$, $d(c\|p)\approx(c-p)^2/(2p(1-p))$, so keeping the cumulative divergence of $n$ queries below $\ln(1/\alpha)$ allows at most $m^*\approx\sqrt{2p(1-p)\,n\ln(1/\alpha)}$ excess on-task queries. For Markov sessions, $\Lambda$ is replaced by the log of the dominant eigenvalue of the tilted transition kernel \citep{chetrite2015nonequilibrium}.

\paragraph{Corollary~\ref{cor:types}.} Given types, the posing accounts' queries have exactly the benign law of their type, so only the count of on-task accounts carries information. It is Binomial$(A,q)$ without the campaign and shifted by $K(1-q)$ with it; the Gaussian approximation gives a divergence of about $K^2(1-q)/(2Aq)$ per week, and Wald's bound over $w$ weeks gives $K^*$.

\paragraph{Proposition~\ref{prop:rand}.} Under the sharp null, $B$'s traffic is the same under every assignment, so $V(Z,Y_B)$ is one draw from the distribution of $V(Z',Y_B)$ over the design \citep{fisher1935design,imbens2015causal}.

\section{The theory in more detail}
\label{app:theory}

\paragraph{Reading the barrier.} Two populations with the same attacker can have very different barriers, because $B_\alpha(A)$ depends only on benign traffic. Every additional benign group raises the bar; a campaign that copies benign behaviour lowers its own score. The propositions are standard arguments, stated because they organise the experiments and say which quantities to measure, not because they are technically deep.

\begin{figure}[ht]
\centering
\includegraphics[width=0.6\linewidth]{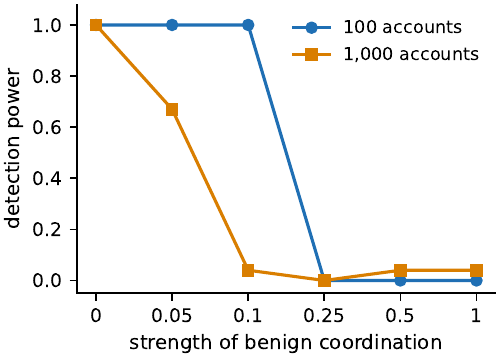}
\caption{The barrier in a controlled setting. Synthetic benign users with topic clusters and trends of increasing strength; campaigns of 10{,}000 extraction queries. A single tight benign cluster per window erases detection power (at strength 0.25 for 100 accounts; at 1{,}000 accounts power falls at 0.05 and vanishes at 0.1). Details in Appendix~\ref{app:controlled}.}
\label{fig:barrier}
\end{figure}

\paragraph{Max-type statistics.} The exact form of Proposition~\ref{prop:max} is $\bar F(\tau)=-\ln(1-\alpha)/\nu$. Figure~\ref{fig:barrier} shows the abrupt collapse it predicts in a controlled study. Our data refute its independent-groups model for the two-scale score, whose threshold stayed flat as populations grew, and match it only qualitatively for $T_{\max}$, whose threshold did rise.

\paragraph{Sequential detection in practice.} Proposition~\ref{prop:seq} changes how to read mimicry: padding that halves a detector's weekly rate multiplies the time a defender must watch. A conformal test martingale replaces the unknown likelihood ratio, but one week can contribute only a bounded amount of evidence, so even an obvious campaign takes about three weeks to flag at $\alpha=1\%$ with our history length (Table~\ref{tab:timefrontier}).

\paragraph{The least-favourable campaign.} Proposition~\ref{prop:lfc} says the least-detectable campaign takes its on-task prompts from what real users already ask. With $p=1.48\%$, twelve weeks of 1{,}640 queries and $\alpha=1\%$, the square-root law allows about 51 extra on-task queries: against an ideal monitor that knows which accounts to watch, covert extraction is limited to the task's benign base rate. Corollary~\ref{cor:types} removes that knowledge; capacity is then counted in accounts, and the bound of about 14 posing accounts over twelve weeks at $A=10^5$ is small because on-task users are rare ($q=0.25\%$ of benign account-weeks), not because the monitor is clever.

\takeaway{What the theory tells a provider.}{Detection is decided by the gap between a campaign and the most similar benign group, not by any signature of extraction. Measure your benign coordination; expect mimicry to cost time rather than to blind you; and expect campaigns built from real users' requests to be the hard case.}

\section{The monitor in detail}
\label{sec:monitor-full}

The monitor answers one question each week: \emph{is there a group of accounts whose queries are jointly unusual, compared with how groups of ordinary users looked in past weeks?} Figure~\ref{fig:monitor} shows the pipeline; this section explains each step and why it is there.

\paragraph{Step 1: a fingerprint for each account.} Every query $x$ is embedded with a sentence encoder $\varphi$ (MiniLM; \citealp{wang2020minilm,reimers2019sbert}). To summarise an account's queries we use random Fourier features \citep{rahimi2007random}: for $m$ random frequency vectors $\omega_j$ drawn at a bandwidth $h$, the account's \emph{amplitude} at frequency $j$ is
\begin{equation}
\rho_a(\omega_j)=\sum_{i=1}^{n_a} e^{-\mathrm{i}\,\omega_j\cdot\varphi(x_i)},
\end{equation}
where $n_a$ is the account's number of queries. Intuitively, each frequency is a ``probe'' of one direction and scale in embedding space; an account that keeps returning to the same region of topic space produces large, aligned amplitudes, and one that wanders produces amplitudes that cancel. A coarse bandwidth ($h=4$) sees broad topics; a fine one ($h=0.5$) sees near-duplicates and templates.

\paragraph{Step 2: compare with normal traffic (whitening).} Benign traffic is far from uniform: some topics are always popular. We remove that by \emph{whitening} each account's amplitudes against the benign traffic of the four preceding weeks,
\begin{equation}
y_a=\Sigma^{-1/2}\,\frac{\rho_a-n_a\mu}{\sqrt{n_a}},
\end{equation}
where $\mu$ and $\Sigma$ are the mean and covariance of one benign query's amplitudes. After whitening, an account that sends independent draws from last month's traffic has a vector $y_a$ with identity covariance, whatever the popularity of the topics it visits. Using a \emph{lagged} reference matters: it absorbs slow drift in what people ask about.

\paragraph{Step 3: look for accounts that move together.} Stack the whitened vectors into a matrix $Y$ (one row per account) and form the cross-account matrix $C=YY^\top/r$, where $r$ is the number of whitened components. If accounts were independent, $C$ would look like a random matrix with a predictable spread of eigenvalues \citep{marchenko1967distribution}. A campaign whose accounts jointly concentrate on a region adds a ``spike'': one eigenvalue rises above the rest, and its eigenvector points at the colluders \citep{baik2005phase}. The detector takes the fewest accounts that carry half of the top eigenvector's weight as its candidate group $G$.

\paragraph{Step 4: score the group.} For a group $G$, the \emph{structure factor} $\Sb_h(G)$ measures how concentrated its pooled queries are at bandwidth $h$ relative to independent draws ($\Sb=1$ for independent queries, larger for concentrated ones). The \emph{two-scale score} compares coarse and fine concentration,
\begin{equation}
T_{\rm 2s}(G)=\log\Sb_{h=4}(G)-\log\Sb_{h=0.5}(G).
\end{equation}
A group that concentrates on a topic without the near-duplicate clumping typical of benign groups scores high. Because the score is a ratio within one group, weekly drift and changes in population size largely cancel, which is why it is the one statistic in our study that stayed calibrated under rolling calibration. (We designed it around deduplication; Section~\ref{sec:results} shows that this is not why it works.)

\paragraph{Step 3$'$: hollowing.} The plain matrix $C$ has a weakness that turned out to matter. Its diagonal holds each account's own concentration $\|y_a\|^2$, so an individually topical benign user---someone who only ever asks about chemistry homework---can dominate its top eigenvector. Padding exploits the same weakness from the other side: mixing natural text into each colluder account shrinks the \emph{size} of each colluder's deviation. The fix is to look only at \emph{directions}. The \emph{centred, hollowed} matrix
\begin{equation}
R_c=\Delta^{-1/2}\bigl(\tilde Y\tilde Y^\top-\Delta\bigr)\Delta^{-1/2},\qquad \Delta=\mathrm{diag}\bigl(\|\tilde y_a\|^2\bigr),
\end{equation}
where $\tilde Y$ is $Y$ with its mean row removed, deletes each account's self-term, normalises every account to unit length and removes the week's common drift. Its top eigenvector picks accounts whose deviations point the same way, however large or small those deviations are. Fifty accounts that share a mathematical core form a correlated block in $R_c$ however much natural text surrounds it. Deleting the diagonal of a Gram matrix is an established device for heteroskedastic data \citep{cai2021subspace,zhang2022heteroskedastic}. We call selection on $R_c$ followed by the two-scale score \emph{hollowed selection}.

\paragraph{Step 3$''$: more than one group.} In large populations several benign groups can out-rank a campaign on $R_c$. \emph{Multi-outlier selection} therefore scores a group from every eigenvector of $R_c$ that rises above the random-matrix bulk (estimated per window by matching the median eigenvalue to the Marchenko--Pastur law; up to 20 eigenvectors, each spread over at least 5 accounts) and keeps the largest score, $T_{\max}$.

\paragraph{Step 5: calibration.} This is where most of the engineering lies. A threshold must reflect benign weeks as they really are: topical users, trends, bulk users, evaluation scripts. We therefore never simulate the null by shuffling accounts or queries. Each threshold is the $(1-\alpha)$ quantile of the same statistic over \emph{whole} historical windows---either the eight preceding weeks (\emph{rolling}) or all prior weeks (\emph{conformal}). Each historical window is whitened against its own lagged reference, exactly as a test window would be. Section~\ref{sec:calib-full} shows why every shortcut fails.

\paragraph{Cost.} The eigenvalues of $C$ come from the $r\times r$ matrix $Y^\top Y$, so the cost of the spectral step does not grow with the number of accounts; forming $Y$ is linear in the number of queries. A weekly window of a few hundred accounts took about 0.2\,s on one T4 GPU and a 1{,}000-account window about 0.5\,s.

\paragraph{Baselines.} We compare against a whole-window test of distribution shift (the \emph{global stream}, $\Sb$ of all accounts together), the MMD \citep{gretton2012kernel} traffic-window test of \citet{liu2026embarrassingly}, a per-account test, Higher Criticism over per-account $p$-values \citep{donoho2004hc}, Fraudar on the account-by-topic matrix \citep{hooi2016fraudar}, a test on the busiest accounts' share of traffic, a dataset-versus-chat classifier and a topic-concentration statistic on the selected group, and a scan of topic-cell volumes against recent history. All share the same lagged references and calibration.

\section{Development-week results in full}
\label{sec:results-full}

This section reports the development weeks (WildChat-1M). Section~\ref{sec:holdout} then re-runs the final pipeline on untouched weeks, and where the two disagree, the untouched weeks decide. Unless stated otherwise, colluders are $K=50$ volume-matched accounts sending about 1{,}640 queries a week, windows are whole weekly populations of a few hundred accounts, and intervals are 95\% bootstrap intervals over test weeks.

\paragraph{A note on uncertainty.} Weeks, not windows, are the independent unit. Zero false alarms in 960 windows would bound the false-alarm rate at 0.3\% only if windows were independent; with 24 weeks as the unit, the bound is about 12\%. Run-to-run variation is also larger than any single interval: the same cell (two-scale detection of MetaMathQA campaigns) gave 0.79, 0.85, 0.88 and 0.93 in four runs that differed in campaign draws and window subsamples. We therefore compare detectors within a run and treat differences below about 0.15 between runs as noise.

\subsection{Calibration is the first thing that breaks}
\label{sec:calib-full}

\begin{figure}[t]
\centering
\includegraphics[width=\linewidth]{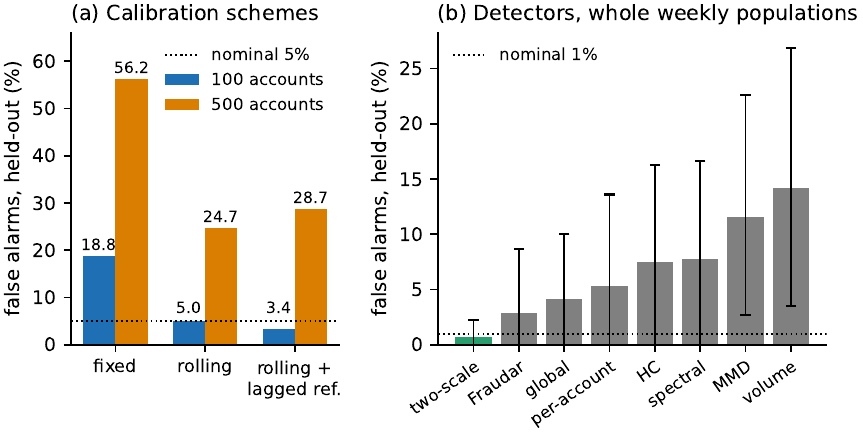}
\caption{Calibration on real traffic. (a) False alarms on held-out weeks for three calibration schemes at a nominal 5\%. Calibrating once on early weeks fails; resampling whole windows from the four preceding weeks, each whitened against its own lagged reference, restores the nominal rate in 100-account windows but not in 500-account windows. (b) Per-detector false alarms at a nominal 1\% over whole weekly populations, with week-clustered intervals: only the two-scale test is calibrated under rolling calibration.}
\label{fig:calib-full}
\end{figure}

A provider's first surprise will be that thresholds do not transfer from one month to the next. Thresholds set on early weeks fired on 12\% of later null weeks against a nominal 1\%: later weeks contain more extreme individual users than any calibration week, and each week's traffic as a whole drifts far from any fixed reference. Resampling \emph{individual accounts} to build a null, the textbook shortcut, also fails, because it destroys the clusters and trends that cross-account tests respond to. What works is resampling \emph{whole historical weeks}, each whitened against its own lagged reference (Figure~\ref{fig:calib-full}a). Even then, over whole weekly populations only the two-scale test stayed calibrated: 0.7\% false alarms at a nominal 1\%, against 2.9\% for Fraudar, 4.1--7.8\% for the global-stream, per-account, Higher Criticism and spectral tests, 11.6\% for MMD and 14.2\% for the volume test (Figure~\ref{fig:calib-full}b). Population sizes vary from week to week, and most statistics move with both size and drift; the two-scale score is a ratio within one group, and both effects largely cancel.

\takeaway{For providers.}{Calibrate on whole historical windows against a lagged reference. A null built by shuffling accounts or queries will look fine in a notebook and fire constantly in production.}

\subsection{What is detectable: coordinated, dataset-like campaigns}
\label{sec:deploy-full}

Table~\ref{tab:deploy} is the core positive result. Campaigns that send prompts from a generated, topically focused dataset were caught most weeks, while no per-account monitor caught anything: volume matching works, and each colluder alone looks ordinary. Broad datasets and real user prompts were not caught---exactly what the barrier predicts for campaigns that look like ordinary users. When the two-scale test fired, the flagged group was almost entirely colluders (precision 1.00 at $K=50$, 0.94--0.96 at $K=10$), but it contained only the 12--31\% of colluders carrying most of the eigenvector's weight. That suits manual review: a handful of accounts that are almost surely colluders, not an enumeration of the campaign.

\paragraph{Its edge is calibration, not ranking.} With all detectors forced to the same realised false-alarm rate by an oracle threshold, the two-scale test caught 0.80 of MetaMathQA campaigns at 1\%, the global stream 0.70 and MMD 0.58; over all operating points, the global stream ranks campaigns slightly \emph{better} (AUC 0.97 against 0.88). The two-scale test's practical advantage is that its threshold transfers across weeks. At stricter oracle false-alarm rates its advantage grows (in a separate run with more null windows: 0.88 at $10^{-3}$ against 0.70 for the global stream and 0.39 for MMD), but the $10^{-3}$ point rests on about three null exceedances and should be read as indicative.

\subsection{What it actually detects: topic shift, not extraction}
\label{sec:mechanism-full}

We designed the two-scale test on the hypothesis that generated distillation sets are concentrated by topic but deduplicated, while benign groups are clumpy. Two ablations, with predictions locked in advance, rejected that story: four rephrasings of each problem cut detection only from 0.88 to 0.80, and exact repeats only from 0.79 to 0.73. What the data support is plainer. Coordinated campaigns of \emph{human-written} GSM8K questions were caught 0.94 of the time---more often than generated MetaMathQA (0.79)---and so were MetaMathQA prompts behind a fixed template (0.94). The test detects \emph{a coordinated group sending prompts concentrated in a topic region that ordinary traffic under-represents}, whatever the reason.

Two consequences matter for a provider. First, simpler source-shift statistics are not enough: a dataset-versus-chat classifier and a topic-concentration statistic applied to the same selected group ranked campaigns moderately (AUC 0.69--0.84) but caught none at a 1\% false-alarm rate; the most dataset-like accounts in every week are benign. Second, the test cannot tell a distiller from a legitimate customer. Benign groups sweeping the GSM8K test set from 3 or 10 accounts triggered it in 79--80\% of weeks, as often as distillation; broad MMLU sweeps did so in only 2--7\%.

\takeaway{For providers.}{Expect evaluation pipelines in the alarms. A firing test is a pointer for review, not a verdict.}

\subsection{Mimicry, and a fix that looks at directions instead of sizes}
\label{sec:mimicry-full}

An attacker who knows a monitor exists has an obvious move: make each colluder look more like a user by mixing natural text into its queries. Mixing half natural prompts into every colluder account, or following every MetaMathQA prompt with a real WildChat follow-up turn, cut two-scale detection from 0.85 to 0.21--0.24 on the development weeks---a fourfold drop, though still well above the false-alarm rate. (On untouched weeks it fell to zero; Section~\ref{sec:holdout}.) We tried four repairs that pool topic cells across accounts---robust per-cell statistics, per-cell lagged scoring with shrinkage, a factor model of cell drift \citep{bai2002determining,ahn2013eigenvalue}, and gating behind the two-scale test. They rank padded campaigns better (AUC up to 0.86) but none reached a usable false-alarm rate at useful power (Appendix~\ref{app:extra}).

The repair that worked came from asking \emph{why} padding works. It dilutes the \emph{size} of each colluder's deviation from normal traffic, but not its \emph{direction}: fifty accounts that each spend half their queries on math still point the same way. The plain matrix $C$ weighs accounts by size, so individually topical benign users dominate it. The centred, hollowed matrix $R_c$ (Section~\ref{sec:monitor}) weighs only directions. With conformal thresholds over all prior weeks, hollowed selection caught MetaMathQA campaigns every week and padded campaigns 0.46--0.48 of the time, against 0.00--0.02 for selection on $C$ under the same calibration, with no false alarm in 960 windows from 24 weeks. Hollowing responds to \emph{many} coordinated accounts: a benign evaluation sweep from ten accounts was flagged 0.97 of the time, one from three accounts 0.42.

Removing the common drift mode also gave a direct reading of the quantity the barrier depends on. Net of each account's own topicality and the week's drift, the top eigenvalue of $R_c$ in real weeks was only 1.14 times that of weeks with queries shuffled across accounts: on these weeks, benign accounts were barely more coordinated than chance. (On the untouched weeks the ratio was 2.22; Section~\ref{sec:holdout}.)

\subsection{Scale: selection fails before the barrier does}
\label{sec:scale-full}

\begin{figure}[t]
\centering
\includegraphics[width=0.62\linewidth]{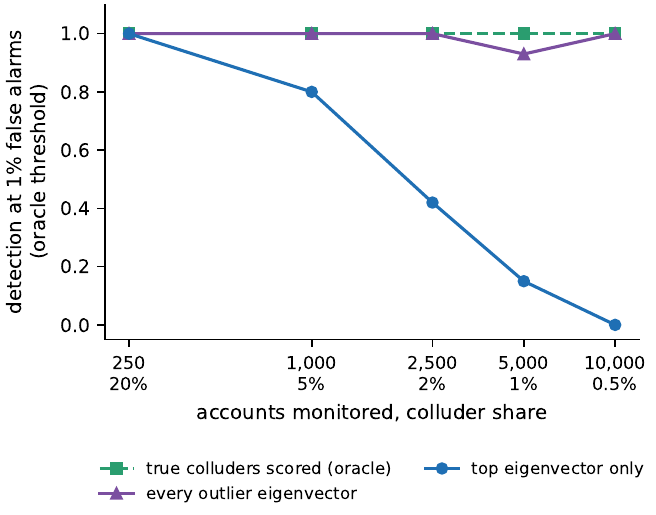}
\caption{Scale, with the campaign fixed at 1{,}640 queries over 50 colluders, in populations pooled from account-weeks across the year (oracle thresholds at 1\% false alarms). The detector's own group selection fails as colluders become rarer, although the true colluder set still scores above the threshold at every size; scoring a group from every outlier eigenvector of $R_c$ closes the gap, but only with oracle thresholds.}
\label{fig:scale-full}
\end{figure}

Production populations hold $10^5$ accounts or more, with colluders far below 1\%. Our weekly populations hold a few hundred. We approached the gap in three ways, and the honest summary is negative for deployment.

\paragraph{Smaller colluder shares.} Holding the campaign fixed at 1{,}640 queries and growing the population from 250 to 1{,}000 account-weeks, oracle detection decayed smoothly with colluder share (0.84, 0.60, 0.35 at 20\%, 10\%, 5\%), as the barrier predicts. Calibration was a separate and worse bottleneck: deployed false alarms in these multi-week populations were 14--90\%. Sharding a population into small windows with one colluder each defeated every detector in all 352 shard tests.

\paragraph{Up to $10^4$ accounts.} In populations pooled from account-weeks across the year, the two-scale test's power fell from 1.00 to 0.00 as colluders went from 20\% to 0.5\% (Figure~\ref{fig:scale-full}). But scoring the \emph{true} colluder set with the same statistic exceeded the threshold at every size: the campaign was still there to be seen; the detector simply stopped picking the right accounts. The barrier did not rise as Proposition~\ref{prop:max} would predict; the null threshold of the two-scale score stayed flat between 2.95 and 4.35. Selecting groups from every outlier eigenvector of $R_c$ closed that gap with oracle thresholds: unpadded campaigns were found in 93--100\% of windows up to $10^4$ accounts (0.5\% colluders), padded ones in 77--100\% up to 5{,}000 accounts and 17\% at $10^4$.

\paragraph{Why this is not yet a production result.} Three caveats bind together. These numbers use oracle thresholds; the best deployed calibration of multi-outlier selection reached 5\% false alarms, not 1\%, at 5{,}000--7{,}552 account-weeks, where it caught unpadded campaigns 0.53--1.00 of the time and padded ones 0.03--0.10. Pooling account-weeks across a year removes most same-week benign coordination, so these populations probably understate the barrier. And sparse scans that look for small dense groups did not help: they were captured by individually heavy benign accounts just as the plain spectrum was.

\takeaway{For providers.}{At realistic colluder shares the bottleneck is first \emph{finding} the right accounts and then \emph{calibrating}. We did not solve either at production shares. Query monitoring is far stronger as a complement to metadata linkage: given 30\% of a campaign's colluders, adding the 50 accounts most correlated with them in $R_c$ recovered 96--100\% of all colluders in populations of 2{,}500--10{,}000 accounts (oracle thresholds and known campaign size).}

\subsection{Attacks on the monitor itself}
\label{sec:attacks-full}

\paragraph{Optimising against the statistic.} An attacker that knows the two-scale statistic, and uses its own benign pool to estimate it, ordered MetaMathQA prompts greedily so that their running coarse fingerprint stayed near zero. Without any padding, this cut weekly detection from 0.80 to 0.31.

\paragraph{Choosing who gets flagged.} Giving the 30\% most distinctive queries to five sacrificial accounts kept detection on $C$ at 0.92, but the flagged group then contained none of the other 45 colluders. Banning flagged accounts does not hurt a campaign that chooses whom the detector flags. Hollowed selection resisted this attack on the untouched weeks (Section~\ref{sec:holdout}).

\paragraph{Teaching the monitor that the campaign is normal.} Lagged calibration has a cost the drift results hide. A campaign that ramps up linearly through the eight calibration weeks contaminates both the lagged reference and the calibration windows; at full volume it was caught 0.12 of the time, against 0.95 for the same campaign arriving abruptly. This is the ``boiling frog'' poisoning known from PCA-based traffic anomaly detection \citep{rubinstein2009antidote}. A reference frozen from early weeks recovered some of it (0.44, at 5.2\% false alarms). On the untouched weeks it sharply cut the detection of every detector (Section~\ref{sec:holdout}).

\section{The controlled study and the spectral calculus}
\label{app:controlled}

\paragraph{Spectral detection condition.} Split a campaign with structure factor $\Sb$ at random across $K$ accounts. Each account has $S\approx1-(1-\Sb)/K$, nearly ordinary, and each colluder pair has cross-correlation $-(1-\Sb)/K$, invisible to pairwise tests once $K$ is large; summed over all ordered pairs these contribute $-(K-1)(1-\Sb)/K$, which together with the self-terms gives back $\Sb$, so the union keeps $\Sb$. In the cross-account matrix this is a rank-one spike of strength $\ell=\Sb-1$, producing an outlier at $(1+\ell)(1+\gamma/\ell)$ when $|\ell|>\sqrt\gamma$, with $\gamma=A/m$ \citep{baik2005phase,benaych2011eigenvalues}. A spectral test therefore needs $|\Sb-1|\sqrt m\gtrsim\sqrt A$, while a test told which accounts collude needs only $|\Sb-1|\sqrt m\gtrsim1$; an exhaustive scan over subsets pays about $\sqrt{2K\ln(eA/K)}$, which is the statistical--computational gap of sparse PCA and planted submatrices \citep{berthet2013optimal,ma2015computational}. Campaigns that cover space \emph{more evenly} than independent sampling ($\Sb<1$) produce an outlier \emph{below} the bulk, a signal pairwise similarity cannot see by construction; on a 2-D torus the measured lower outlier matched the BBP prediction to four decimals, and over 20{,}000 null worlds analytic Tracy--Widom thresholds \citep{tracy1994level,bornemann2010numerical} matched false-alarm rates from $10^{-1}$ to $10^{-4}$. No real extraction strategy we tested triggered the lower edge, so it plays no role in the main text.

\paragraph{CIFAR-10 regime map.} Against a ResNet-18 victim \citep{he2016resnet} on CIFAR-10 \citep{krizhevsky2009cifar}, every active extraction strategy concentrated in the victim's embedding and was detected at the upper edge. The spectral test dominated when many colluders sent few queries each and $K\lesssim2.3\sqrt A$; per-account tests and Higher Criticism sufficed with few heavy colluders; the global stream won with many colluders.

\paragraph{The dose--response study.} Benign users with lognormal volumes, topic clusters of 10 accounts sending half their queries to one narrow region, and a weekly trend followed by a fifth of accounts were added with strength $s$ (Table~\ref{tabA:dose}). Only resampling whole historical windows calibrated the suite (0.6\% false alarms at $s=1$; independent-account resampling left the spectral test at 31\% and the global test at 100\%). One tight benign cluster per window erased geometric power: benign ten-account clusters had union $\Sb$ of 83--159, against 26--30 for the attacks. A task-aware suite on the victim's low-margin queries tolerated one cluster at 100 accounts and a few at 1{,}000, then failed.

\begin{table}[h]
\centering
\caption{Power against 10{,}000-query campaigns as benign structure strength $s$ grows (24 windows per cell, window-resampled calibration). Geometry suite: spectral, Higher Criticism and global tests on the victim embedding. Task-aware suite: the same aggregations applied to each account's share of low-margin queries.}
\label{tabA:dose}
\small
\begin{tabular}{lllcccccc}
\toprule
$A$, $K$ & Strategy & Suite & $s=0$ & 0.05 & 0.1 & 0.25 & 0.5 & 1 \\
\midrule
100, 20 & entropy & geometry & 1.00 & 1.00 & 1.00 & 0.00 & 0.00 & 0.00 \\
100, 20 & entropy & task-aware & 1.00 & 1.00 & 1.00 & 1.00 & 0.12 & 0.17 \\
100, 20 & $k$-center & geometry & 1.00 & 1.00 & 1.00 & 0.00 & 0.00 & 0.00 \\
100, 20 & $k$-center & task-aware & 1.00 & 1.00 & 1.00 & 0.79 & 0.04 & 0.08 \\
1000, 100 & entropy & geometry & 1.00 & 0.67 & 0.04 & 0.00 & 0.04 & 0.04 \\
1000, 100 & entropy & task-aware & 1.00 & 0.96 & 0.67 & 0.33 & 0.04 & 0.04 \\
1000, 100 & $k$-center & geometry & 1.00 & 0.38 & 0.04 & 0.00 & 0.04 & 0.04 \\
1000, 100 & $k$-center & task-aware & 1.00 & 0.21 & 0.25 & 0.08 & 0.04 & 0.04 \\
\bottomrule
\end{tabular}
\end{table}

\section{Additional real-traffic results}
\label{app:extra}

\paragraph{Small monitored populations.} Table~\ref{tabA:smallpop} reports the first real-traffic protocol, with equal-volume colluders forming 46--94\% of the window. A narrow capability campaign of 5{,}000 prompts was caught 94\% of the time in 100-account windows; broad campaigns only when they formed almost all of the traffic. In a separate run that varied the embedding dimension (not shown in the table), keeping more of the embedding helped weaker signals: from 32 to 384 dimensions, broad 5{,}000-prompt campaigns rose from 0.18 to 0.35 and padded 20{,}000-prompt campaigns from 0.36 to 0.60.

\begin{table}[h]
\centering
\caption{Paired detection in small monitored populations (8 held-out weeks, 3 campaigns $\times$ 3 windows per week; $\alpha=5\%$). Share is the campaign's fraction of window traffic.}
\label{tabA:smallpop}
\footnotesize\setlength{\tabcolsep}{3pt}
\begin{tabular}{llrrccc}
\toprule
$A$, $K$ & Campaign & $N$ & Share & Fixed & Rolling & Rolling + lag \\
\midrule
100, 20 & random natural & 20{,}000 & 0.94 & 0.00 & 0.00 & 0.00 \\
100, 20 & broad capability (1 of 10) & 5{,}000 & 0.81 & 0.03 & 0.04 & 0.21 \\
100, 20 & broad capability & 20{,}000 & 0.94 & 0.82 & 0.90 & 0.93 \\
100, 20 & narrow capability (1 of 30) & 1{,}000 & 0.46 & 0.00 & 0.00 & 0.03 \\
100, 20 & narrow capability & 5{,}000 & 0.81 & 0.78 & 0.93 & \textbf{0.94} \\
100, 20 & broad, 50\% natural padding & 20{,}000 & 0.94 & 0.25 & 0.04 & 0.38 \\
500, 100 & broad capability & 5{,}000 & 0.46 & 0.10 & 0.07 & 0.29 \\
500, 100 & narrow capability & 5{,}000 & 0.46 & 0.44 & 0.78 & 0.71 \\
\midrule
\multicolumn{4}{l}{Held-out false alarms at $A=100$ / 500} & 18.8\% / 56.2\% & 5.0\% / 24.7\% & 3.4\% / 28.7\% \\
\bottomrule
\end{tabular}
\end{table}

\paragraph{Oracle comparison of statistics.} Table~\ref{tabA:oracle} compares detectors at matched realised false-alarm rates. With 96 null windows the 1\% threshold is set by about one null window, so these values are fragile.

\begin{table}[h]
\centering
\caption{Primary comparison: detection at oracle-matched realised false-alarm rates (thresholds from the pooled held-out null windows, which a deployed system cannot set), with AUC. Weekly populations, $K=50$ volume-matched colluders, about 1{,}640 queries, 24 test weeks. QC is the repaired query-cluster detector (per-cell lagged scoring with shrinkage; see ``Query-level pooling'' below). Bold: best AUC and best TPR at 1\% in each row.}
\label{tabA:oracle}
\footnotesize
\setlength{\tabcolsep}{4pt}
\begin{tabular}{lcccc}
\toprule
& \multicolumn{4}{c}{AUC / TPR at 1\% / TPR at 5\%} \\
\cmidrule(lr){2-5}
Campaign & Two-scale & Global stream & MMD (Liu et al.) & QC \\
\midrule
MetaMathQA (generated) & 0.88 / \textbf{0.80} / 0.81 & \textbf{0.97} / 0.70 / 0.74 & 0.94 / 0.58 / 0.62 & 0.94 / 0.64 / --- \\
Human GSM8K questions & 0.96 / \textbf{0.95} / 0.95 & 0.99 / 0.88 / 0.94 & 0.97 / 0.73 / 0.80 & \textbf{1.00} / 0.93 / --- \\
Half natural, mixed into every account & 0.68 / 0.24 / 0.39 & 0.75 / 0.05 / 0.11 & 0.71 / 0.07 / 0.14 & \textbf{0.86} / \textbf{0.31} / --- \\
Real follow-up after every prompt & 0.68 / 0.24 / 0.40 & 0.75 / 0.06 / 0.11 & 0.72 / 0.11 / 0.20 & \textbf{0.85} / \textbf{0.27} / --- \\
Random natural prompts & 0.48 / 0.00 / 0.07 & 0.39 / 0.00 / 0.04 & 0.39 / 0.02 / 0.02 & 0.52 / 0.10 / --- \\
\bottomrule
\end{tabular}
\end{table}

\paragraph{Mechanism probes.} Table~\ref{tabA:mech}: rephrased variants barely weaken the two-scale test and deduplicating natural prompts does not create a signal. A manipulation check showed that rephrasings lie a median 0.67 apart in the reduced embedding, so only 35\% fall within the fine bandwidth; removing benign follow-up turns moved benign clumping away from that of colluders. Both designed mechanisms were rejected.

\begin{table}[h]
\centering
\caption{Mechanism probes (weekly populations, $K=50$, volume-matched). Rephrased variants barely weaken the test, and deduplicating natural prompts does not create a signal.}
\label{tabA:mech}
\small
\begin{tabular}{lccc}
\toprule
Campaign & Two-scale & Global & MMD \\
\midrule
MetaMathQA, one prompt per source problem & 0.88 & 0.68 & 0.57 \\
MetaMathQA, four variants per source problem & 0.80 & 0.71 & 0.60 \\
WildChat broad cluster, random prompts & 0.01 & 0.01 & 0.07 \\
WildChat broad cluster, farthest-point deduplicated & 0.04 & 0.02 & 0.07 \\
\bottomrule
\end{tabular}
\end{table}

\paragraph{Query-level pooling.} A detector that scores each topic cell's two-scale statistic pooled across all accounts ranks padded campaigns best (AUC 0.82--0.86 against at most 0.75 for account-level detectors under mimicry), but it never reached a usable false-alarm rate at useful power: robust per-cell calibration fired on 28\% of null weeks; per-cell lagged scoring with shrinkage on 15\%; gating behind the two-scale test on 12\%; a factor model of cell drift on 23\%, because two factors carried only 31\% of the variance. A full-history conformal threshold reached calibration, at which point it caught padded campaigns 0.20 of the time---no better than the two-scale test. A trivial per-cell volume scan fired on 12.8\% of null weeks at a nominal 1\% on the development weeks and caught MetaMathQA campaigns 0.36 of the time.

\paragraph{Hollowing.} Table~\ref{tabA:hollow} shows that selection on the uncentred hollowed matrix $R$ ranks padded campaigns almost perfectly at oracle thresholds but was not calibrated (8.4\% false alarms), because the week's common drift dominated it; centring fixed this. Table~\ref{tabA:hollow2} gives the centred version with conformal thresholds and in pooled populations.

\begin{table}[h]
\centering
\caption{Selection on the hollowed matrix $R=\Delta^{-1/2}(YY^\top-\Delta)\Delta^{-1/2}$ (as $R_c$ but without centring) versus $C$, both scored with the two-scale statistic. Oracle TPR at a matched 1\% false-alarm rate (AUC in parentheses); weekly populations, $K=50$ volume-matched colluders. Deployed false alarms: 0.8\% for $C$, 8.4\% for $R$, 7.6\% for the core variant.}
\label{tabA:hollow}
\small
\begin{tabular}{lccc}
\toprule
Campaign & Selection on $C$ & Selection on $R$ & $R$ on topic cores \\
\midrule
MetaMathQA & 0.86 (0.92) & 1.00 (1.00) & 1.00 (1.00) \\
Half natural, mixed & 0.21 (0.66) & 0.94 (1.00) & 0.95 (1.00) \\
Real follow-up after every prompt & 0.23 (0.66) & 0.93 (1.00) & 0.93 (1.00) \\
Human GSM8K questions & 0.93 (0.97) & 1.00 (1.00) & 1.00 (1.00) \\
Random natural prompts & 0.02 (0.52) & 0.00 (0.51) & 0.00 (0.48) \\
GSM8K evaluation sweep, 3 accounts & 0.75 (0.87) & 0.34 (0.61) & 0.02 (0.54) \\
\bottomrule
\end{tabular}
\end{table}
\begin{table}[h]
\centering
\caption{Selection on the centred hollowed matrix $R_c$ against $C$. Top: weekly populations, conformal thresholds over all prior weeks (no false alarm in 960 held-out windows from 24 weeks for either; week-level 95\% upper bound about 12\%). Bottom: pooled populations with the campaign fixed at 1{,}640 queries over 50 colluders, oracle TPR at 1\%, and the share of colluders in the group $R_c$ selects. The $C$ row repeats cells of Table~\ref{tabA:scale10k} from a separate run (0.28 against 0.42 at $A=2{,}500$), a difference within run-to-run variation.}
\label{tabA:hollow2}
\small
\begin{tabular}{lcccccc}
\toprule
\multicolumn{7}{l}{\emph{Weekly, deployed (conformal)}} \\
& MetaMathQA & Mixed & Follow-up & Random & Sweep, 3 accts & Sweep, 10 accts \\
Selection on $C$ & 0.82 & 0.00 & 0.02 & 0.00 & 0.80 & 0.82 \\
Selection on $R_c$ & 1.00 & 0.48 & 0.46 & 0.00 & 0.42 & 0.97 \\
\midrule
\multicolumn{7}{l}{\emph{Pooled, oracle TPR at 1\%}} \\
& $A=250$ & 1{,}000 & 2{,}500 & 5{,}000 & 10{,}000 & \\
$C$, unpadded & 0.97 & 0.75 & 0.28 & 0.12 & 0.05 & \\
$R_c$, unpadded & 1.00 & 1.00 & 0.00 & 0.00 & 0.03 & \\
$R_c$, mixed padding & 1.00 & 0.10 & 0.00 & 0.00 & 0.03 & \\
Colluders in $R_c$'s group & 1.00 & 1.00 & 0.00 & 0.00 & 0.00 & \\
\bottomrule
\end{tabular}
\end{table}

\paragraph{Scale.} Table~\ref{tabA:scale10k} gives the pooled populations with the campaign fixed, and Table~\ref{tabA:multi} multi-outlier selection with its diagnostics. The colluders' eigenvector was always far above the random-matrix bulk (pairwise cosine 0.49--0.61 against a requirement of at most 0.082), but at rank 1--8 rather than first: benign coordinated groups out-ranked it. Raising the number of whitened components from 600 to 1{,}500 at 2{,}500 accounts lifted top-eigenvector selection from 0.00 to 0.57 by changing the ranking. The number of retained outliers reached the cap of 20 from 1{,}000 accounts upward. Sparse scans (truncated power iteration with up to 100 iterations and 20 restarts, and a greedy forward scan) caught 0--4\% of campaigns at every colluder fraction from 1\% to 28\%, while an oracle told the colluders caught all of them.

\begin{table}[h]
\centering
\caption{Scale with the campaign fixed (1{,}640 MetaMathQA queries over $K=50$ colluders) in pooled populations of benign account-weeks, oracle thresholds. ``True colluder set'' scores the campaign's own accounts with the two-scale statistic against the same null threshold.}
\label{tabA:scale10k}
\small
\begin{tabular}{rccccc}
\toprule
$A$ & Colluder share & Null median & Null 99\% & Detector TPR at 1\% & True colluder set above threshold \\
\midrule
250 & 20\% & 1.27 & 3.28 & 1.00 & 1.00 \\
1{,}000 & 5\% & 1.21 & 3.16 & 0.80 & 1.00 \\
2{,}500 & 2\% & 1.05 & 4.35 & 0.42 & 1.00 \\
5{,}000 & 1\% & 0.94 & 3.18 & 0.15 & 1.00 \\
10{,}000 & 0.5\% & 0.61 & 2.95 & 0.00 & 1.00 \\
\bottomrule
\end{tabular}
\end{table}
\begin{table}[h]
\centering
\caption{Multi-outlier selection at scale: oracle TPR at a matched 1\% false-alarm rate in pooled populations, campaign fixed at 1{,}640 queries over 50 colluders. Diagnostics: rank of the eigenvector of $R_c$ most aligned with the colluders (0 = top eigenvector), and their mean pairwise cosine $\rho$ against the BBP requirement $\sqrt{A/r}/K$.}
\label{tabA:multi}
\small
\begin{tabular}{rccccccc}
\toprule
& & \multicolumn{2}{c}{Top eigenvector} & \multicolumn{2}{c}{$T_{\max}$} & & \\
\cmidrule(lr){3-4}\cmidrule(lr){5-6}
$A$ & Colluders & unpadded & padded & unpadded & padded & Colluder rank (unp./pad.) & $\rho$ vs needed \\
\midrule
250 & 20\% & 1.00 & 1.00 & 1.00 & 1.00 & 0 / 0 & 0.49 vs 0.013 \\
1{,}000 & 5\% & 1.00 & 0.07 & 1.00 & 1.00 & 0 / 1 & 0.58 vs 0.026 \\
2{,}500 & 2\% & 0.00 & 0.00 & 1.00 & 0.83 & 1 / 2 & 0.58 vs 0.041 \\
5{,}000 & 1\% & 0.00 & 0.00 & 0.93 & 0.77 & 2 / 3 & 0.60 vs 0.058 \\
10{,}000 & 0.5\% & 0.03 & 0.03 & 1.00 & 0.17 & 3 / 8 & 0.61 vs 0.082 \\
\bottomrule
\end{tabular}
\end{table}

\paragraph{Statistics that failed.} A structure-factor profile over six bandwidths scored against benign groups of equal coarse concentration detected nothing under padding; a core-plus-halo statistic was at chance (AUC 0.48); a behavioural prefilter on the predictability of each account's topic sequence kept none of the colluders, because a campaign confined to a few topics is \emph{more} predictable than a human; its two-sided version kept 13\% of colluders and every benign evaluation account. A dual-reference test that tracks the growing gap between scores against a frozen and a lagged reference (in the spirit of early-warning signals of regime shifts; \citealp{scheffer2009early}) detected ramps 0.03 of the time. Mixing a fraction $f$ of natural prompts into every colluder account moved oracle two-scale detection from 0.81 ($f=0$) to 0.00 ($f=0.75$), crossing 0.5 at a campaign-to-reference MMD$^2$ of about 0.035, while the most distinctive benign groups were much farther from the reference (median 0.145) and were not flagged: distance from the reference is not the gap that matters.

\section{The capability--detectability frontier in full}
\label{sec:frontier-full}

Detection rates alone do not say whether a defence matters. An attacker who evades the monitor but learns nothing has not won. To weigh the two, we distilled every campaign we studied into a small student and measured what it learned.

\paragraph{Protocol.} Each campaign is turned into exactly the (prompt, answer) pairs it would collect, and a student is fine-tuned on them. For dataset prompts the answers are the dataset's; for natural prompts they are the ChatGPT replies recorded in WildChat. Capability is accuracy on GSM8K test problems \citep{cobbe2021gsm8k}. Answers come from datasets and logs, not from a live victim; Section~\ref{sec:limitations} discusses what that leaves open.

\subsection{Two regimes: elicitation and extraction}

The first measurements showed no cost of mimicry at all, and the reason turned out to be informative. A Qwen2.5-0.5B student \citep{qwen2024qwen25} gained 9.8 GSM8K points from only 100 math prompts and no more from 900; mixing in natural prompts, or restricting to natural-looking math prompts, gained as much. A larger student on a harder target (Qwen2.5-1.5B on MATH-500) behaved the same way: flat gains from 1{,}000 to 4{,}000 queries. These students already had the skill; a hundred examples teach them the answer format. We call this the \emph{elicitation} regime, in line with the finding that imitation of proprietary models transfers style far more than capability \citep{gudibande2024false}. In this regime stealth is free, and the attacker can also stay small.

\begin{figure}[t]
\centering
\includegraphics[width=0.85\linewidth]{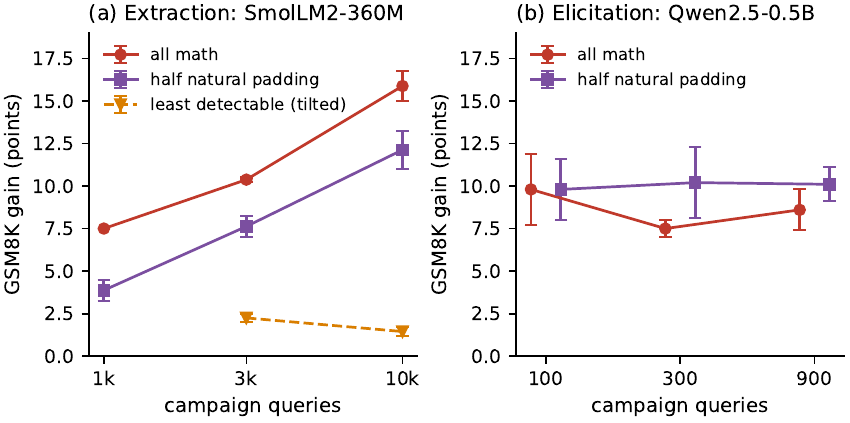}
\caption{The price of stealth depends on whether the student already has the skill. (a) Extraction regime: SmolLM2-360M, which starts at 3.5\% on GSM8K, keeps gaining with budget; padding half the campaign costs 3--4 points at every budget, and the least-detectable (tilted) campaign gains almost nothing (bars: range over two seeds). (b) Elicitation regime: Qwen2.5-0.5B on GSM8K (three seeds) saturates within 100 queries, and padding costs nothing measurable.}
\label{fig:stealth-full}
\end{figure}

A price of stealth can only appear when the student \emph{lacks} the capability, so that every useful query counts. SmolLM2-360M \citep{allal2025smollm2} solves only 3.5\% of GSM8K. Fully fine-tuned on MetaMathQA math prompts, it gained 7.5, 10.4 and 15.9 points at 1{,}000, 3{,}000 and 10{,}000 queries---the \emph{extraction} regime (Figure~\ref{fig:stealth-full}a). Here padding half of every campaign with natural prompts cost 3.6, 2.8 and 3.8 points, or 48\%, 27\% and 24\% of the gain, consistently across seeds. The natural prompts contributed nothing: a padded campaign of $n$ queries learned about as much as an unpadded campaign of $n/2$. The cost of mimicry is the queries spent on it.

\subsection{The covert end: campaigns that evade every detector}

Proposition~\ref{prop:lfc} says what the least-detectable campaign looks like: real benign prompts, re-weighted toward the useful region. We built it from the attacker's pool---half real WildChat prompts that a classifier places in the math region, half the rest. It evaded every detector (weekly detection at most 0.01 on development weeks, at most 0.04 on untouched weeks) and gained 2.2 points at 3{,}000 queries, against 10.4 for the dataset campaign: covertness cost 79\% of the gain.

An obvious objection is that WildChat's math questions are simply the wrong kind of math. We tested the attacker that decides the question: one that \emph{writes} its own useful prompts in the voice of real users. An instruction-tuned model (Qwen2.5-1.5B-Instruct) rewrote GSM8K-style problems as messages from WildChat users, shown real user messages as style examples and keeping each problem's answer. It could not write like users: a classifier told its rewrites from real prompts with AUC 0.98, and after two more rounds of regenerating every rewrite it caught, 0.97. Only 395 of 5{,}000 rewrites (7.9\%) passed as user-like. We then gave the attacker the best we could: resampling all rewrites to match the distribution of real on-task prompts as closely as the classifier could measure. That campaign evaded every detector and gained 4.4 points---42\% of the dataset campaign's gain, the most any covert campaign achieved. Its effective sample was only 151 distinct rewrites.

\takeaway{The frontier, in one sentence.}{In our testbed, every campaign that evaded every detector also gave up 58\% or more of what a detectable campaign transferred, and each was limited by its supply of prompts that are both useful and indistinguishable from users'.}

This conclusion is narrower than it sounds, and we want to be explicit about how narrow. It rests on one 360M student, one benchmark, answers taken from datasets rather than a live victim, and two seeds; the student solves 14 of 400 problems at baseline, so the covert gains of 1.1--4.4 points are differences of 4--18 problems. The 213 rewrites that passed the first filter gained 3.5 points on their own, which may be format learning even in this regime. And the supply limit is exactly what attackers with a user base do not face (Section~\ref{sec:world}). A stronger writer, a live teacher and more seeds are the obvious next tests.

\subsection{Capability held at first detection}

What an attacker optimises is the capability it holds when it is first caught. Table~\ref{tab:timefrontier} combines the gain curves with detection times at the weekly campaign size of our detection experiments (1{,}640 queries a week), under two detection rules applied the same way to every detector. The \emph{first weekly alarm} at a 1\% threshold is fast but accumulates false alarms (null runs alarmed within 16 weeks a quarter of the time on untouched weeks). The \emph{anytime-valid} conformal test martingale holds the probability of any false alarm, ever, at 1\%.

Two lessons stand out. Against the two-scale and multi-outlier tests, padding on the untouched weeks bought evasion for at least a quarter of a year in most runs at a cost of a quarter to a half of the gain. Against hollowed selection it bought nothing: padded campaigns were caught in the first week. The time to an anytime-valid alarm for hollowed selection is the most important number we did not measure.

\section{Distillation details}
\label{app:distill}

Table~\ref{tabA:stealth} gives the first distillation measurement (Qwen2.5-0.5B, LoRA; \citealp{hu2022lora}), Table~\ref{tabA:stealth2} the elicitation regime with three seeds, and Table~\ref{tabA:extract} the extraction regime. The least-detectable (tilted) campaign at 10{,}000 queries is confounded: the attacker's pool held only 2{,}933 on-task prompts, so that campaign was 29\% on task and reused prompts. Defining the task region by value to the target (a GSM8K-versus-chat classifier) left only 972 on-task prompts, and a tilted campaign built from them gained 1.1 points at 3{,}000 queries. A persona campaign, in which each colluder took over a real on-task pool user's week with its on-task queries replaced by rewrites, evaded every query-level detector and gained 3.5 points from 1{,}891 queries; a detector tracking the share of majority-on-task accounts caught it 0.58 of the time.

\begin{table}[h]
\centering
\caption{Detection against what the campaign extracts. Detection: paired power at 1\% (weekly populations, $K=50$, 1{,}641 queries). Distillation: Qwen2.5-0.5B (base) fine-tuned with LoRA on the campaign's (prompt, answer) pairs, GSM8K accuracy on the first 500 test problems (one seed; standard error about 2.2 points). Base model: 34.4\%.}
\label{tabA:stealth}
\small
\setlength{\tabcolsep}{4pt}
\begin{tabular}{lccccc}
\toprule
& \multicolumn{3}{c}{Detection at 1{,}641 queries} & \multicolumn{2}{c}{GSM8K gain (points)} \\
\cmidrule(lr){2-4}\cmidrule(lr){5-6}
Campaign & Two-scale & Global & MMD & 1{,}641 queries & 7{,}550 queries \\
\midrule
MetaMathQA, one prompt per source problem & 0.85 [0.74, 0.94] & 0.65 & 0.56 & +11.4 & +12.6 \\
Half natural prompts mixed into every account & 0.21 [0.10, 0.37] & 0.12 & 0.08 & +12.0 & +11.0 \\
Each prompt followed by a real follow-up turn & 0.24 [0.11, 0.38] & 0.12 & 0.08 & +13.4 & --- \\
Natural WildChat cluster & 0.01 & 0.01 & 0.07 & --- & +6.0 \\
\bottomrule
\end{tabular}
\end{table}
\begin{table}[h]
\centering
\caption{GSM8K gain over the base model (points; mean $\pm$ standard deviation over three seeds) for campaigns of equal query budget. All models are evaluated on the same 400 problems.}
\label{tabA:stealth2}
\small
\begin{tabular}{lccc}
\toprule
Queries & All math & Half natural, mixed & Camouflaged (natural-looking math) \\
\midrule
100 & $+9.8\pm2.1$ & $+9.8\pm1.8$ & $+9.3\pm0.7$ \\
300 & $+7.5\pm0.5$ & $+10.2\pm2.1$ & $+8.7\pm2.3$ \\
900 & $+8.6\pm1.2$ & $+10.1\pm1.0$ & $+11.2\pm1.9$ \\
\bottomrule
\end{tabular}
\end{table}
\begin{table}[h]
\centering
\caption{The extraction regime: gain over the base model (points) for campaigns of equal query budget. SmolLM2-360M (base 3.5\% on GSM8K, 400 problems), full fine-tuning, mean $\pm$ standard deviation over two seeds; Qwen2.5-1.5B with LoRA on MATH-500 (base 22.8\%), one seed. Tilted rows: half on-task WildChat prompts and half off-task (Proposition~\ref{prop:lfc}); the cost is against the all-math campaign of the same size.}
\label{tabA:extract}
\small
\begin{tabular}{lcccc}
\toprule
& Queries & All math & Half natural, mixed & Paired cost \\
\midrule
SmolLM2-360M, GSM8K & 1{,}000 & $+7.5\pm0.0$ & $+3.9\pm0.6$ & $-3.6$ \\
 & 3{,}000 & $+10.4\pm0.1$ & $+7.6\pm0.6$ & $-2.8$ \\
 & 10{,}000 & $+15.9\pm0.9$ & $+12.1\pm1.1$ & $-3.8$ \\
\midrule
SmolLM2-360M, tilted (least detectable) & 3{,}000 & --- & $+2.2\pm0.2$ & $-8.2$ \\
 & 10{,}000 & --- & $+1.5\pm0.2$ & $-14.4$ \\
\midrule
Qwen2.5-1.5B, MATH-500 & 1{,}000 & $+12.0$ & $+11.4$ & $-0.6$ \\
 & 4{,}000 & $+11.6$ & $+9.6$ & $-2.0$ \\
\bottomrule
\end{tabular}
\end{table}

\section{The untouched weeks in full}
\label{sec:holdout-full}

Every choice behind hollowed and multi-outlier selection---hollowing, centring, conformal thresholds over the full history, the bulk-edge factor of 1.05, the minimum group spread of 5 accounts and the cap of 20 eigenvectors---was made after seeing results on the same 24 development weeks. Predictions were locked before each run and many failed, so the choices were not tuned until predictions passed; but each was informed by earlier results on those weeks. Results obtained this way are optimistic by an amount the same data cannot estimate. We therefore froze the whole pipeline and ran it once, with 13 predictions locked beforehand, on data no choice had touched.

\paragraph{Design.} WildChat-4.8M conversations from 6 May to 1 September 2024: 17 test weeks after WildChat-1M ends, with the preceding 31 weeks used only as calibration history (757{,}000 English user turns from 76{,}000 hashed IPs in all). The PCA basis, topics, task-region classifier and every threshold were fitted on history weeks only. Windows held 64--252 accounts, within the development-week range, and each campaign window was paired with the identical window without it. We repeated everything with a second sentence encoder, E5-small-v2---a close relative of MiniLM, from which it is initialised, so not an independent representation---and in whole weekly populations of all accounts with at least 5 queries.

\paragraph{What held.} Calibration held, though not perfectly: conformal thresholds kept the three main detectors at 1.8--1.9\% false alarms against a nominal 1\% with MiniLM (two or three of 17 weeks had any false alarm) and at 0--1.3\% with E5. A provider should budget for about twice the nominal rate. Dataset campaigns were caught 0.86--1.00 of the time in both encoders, and the accounts flagged by multi-outlier selection were colluders 94\% of the time. The least-detectable campaign and random real prompts stayed at or below 0.04 for the three spectral detectors. Most importantly, \textbf{hollowed selection still saw through padding}, more clearly than before: 0.50--0.57 of padded campaigns across three campaign seeds, 0.89 with E5 and 0.96 in whole weekly populations. Run-to-run spread across campaign seeds was at most 0.07, and adding colluders instead of substituting them for benign accounts changed nothing.

\paragraph{What did not.} Under padding, the two-scale test fell from 0.21 to 0.00 and multi-outlier selection from 0.36--0.42 to 0.00, in both encoders and both regimes. Multi-outlier selection pays for its maximum over up to 20 groups inside its threshold, and on these weeks the colluders' group no longer out-scored the most distinctive benign groups---even though the hollowed top eigenvector still found it. The statistic that closed the selection gap at scale with oracle thresholds therefore costs power at a fixed false-alarm rate in weekly populations; a provider should run it beside, not instead of, hollowed selection. Benign coordination was also higher than on the development weeks: the real-to-shuffled ratio of the top eigenvalue of $R_c$ was 2.22 with MiniLM and 3.06 with E5, against 1.14. The barrier's key quantity changes over time, which is why it must be measured, not assumed. The volume baseline with full-history thresholds was calibrated but nearly powerless (0.03 on dataset campaigns); in its rolling form it caught padded campaigns more often than the two-scale test (0.37 against 0.00) in small windows, but fired on 6--11\% of null windows elsewhere.

\paragraph{Attacks on hollowed selection.} Sacrificial scheduling, which hid every other colluder from selection on $C$ (again 0.00 here), did not steer hollowed or multi-outlier selection: both still caught every campaign and flagged 49--59\% of the non-sacrificial colluders, as many as under random assignment. Hollowing normalises each account, so concentrating the most distinctive queries on a few accounts does not move the direction the colluders share. The calibration ramp, by contrast, was far worse than we predicted. Ramping the campaign through the eight calibration weeks cut multi-outlier detection from 0.89 to 0.00, the conformal two-scale test from 0.80 to 0.02, hollowed selection from 0.98 to 0.32 and the rolling two-scale test from 0.79 to 0.32. A long history does not dilute poisoning: a 1\% conformal threshold is the 99th percentile of past null scores, so once contaminated windows exceed 1\% of the history they set it.

\paragraph{Covert is not the same as invisible.} Hollowed selection ranked the least-detectable campaign well above benign windows (AUC 0.97 with MiniLM, 0.92 with E5) without detecting it at deployed thresholds. Random real prompts also ranked above benign windows (AUC 0.88), so part of this signal comes from how we build colluder accounts: an account that sends random draws from pooled traffic is less topical than any real user. The covert campaign may therefore be easier to catch with a sequential test on the hollowed score, or harder once colluders are built from whole real user-weeks. Both are open.

\paragraph{What could not be tested.} We had planned to test real same-week populations far larger than the development weeks. In this period, WildChat-4.8M's weekly populations held only 260--469 accounts with at least 5 queries, so whether real populations of $10^4$ accounts are harder than our pooled ones remains open. At the false-alarm rate measured in whole weeks, multi-outlier selection would flag about 160 benign accounts a week in a population of $10^5$.

\takeaway{What the untouched weeks changed.}{Hollowed selection is the most robust deployed detector we built: near-nominal false alarms, unaffected by sacrificial scheduling, and the only one that still catches padded campaigns at near-nominal false alarms in every setting. Benign coordination varies over time and must be measured. Calibration poisoning by a slow ramp is the most serious threat we found to every detector.}

\section{Real campaigns, relays and randomised interventions}
\label{sec:world-full}

\paragraph{Documented campaigns.} The campaigns reported in 2026 occupy two corners of the frontier, neither of them the covert, self-written corner we studied in most detail. At one corner are high-volume campaigns. The largest Anthropic has reported ran more than 151 million exchanges between May and July 2026 through more than 3{,}500 accounts, peaking near 3 million a day---several hundred exchanges per account per day---and its accounts were linked because they all used one fixed prompt \citep{anthropic2026threat}. Campaigns like this are visible to per-account volume monitors and to template coordination, which our detectors caught almost every week (Section~\ref{sec:mechanism-full}). At the other corner are \emph{relays}: one forwarded almost 300{,}000 customer requests in ten days through 5{,}380 accounts, about six per account per day; another relayed more than 12 million exchanges in fourteen days \citep{anthropic2026threat}. Relays have low per-account volume and on-task content drawn from real users.

\paragraph{Why relays are the hard case.} Our covert campaigns were limited by supply: an attacker without a user base could not obtain or write enough prompts that are both useful and user-like. A relay has an unlimited supply of exactly such prompts. If a relaying account's traffic has the same joint law as the users it relays, Proposition~\ref{prop:law} leaves no query-only monitor able to separate it. That ``if'' is an assumption, not a measurement. A relay that interleaves many users' sessions within one account, or forwards only high-value requests, may differ in session coherence or in how topical each account is, and a monitor could in principle see that. We did not simulate relays; this is the most important experiment we leave open.

\paragraph{Beyond passive monitoring: the provider's own randomness.} Every threshold in this paper is learned from history, and history drifts and can be poisoned. A provider has one resource no attacker controls: its own randomness. If it throttles a flagged group in randomly chosen time blocks, it can test whether the work fails over to other accounts, with a null distribution that follows from the randomisation alone.

\begin{proposition}[Randomisation null]\label{prop:rand}
Fix, before any intervention, a flagged group $G$, a disjoint set $B$ of accounts to watch, and time blocks $1,\dots,L$. Draw a throttle schedule $Z$ for $G$ from a known design, independently of all traffic. For any statistic $V(Z,Y_B)$ of $B$'s traffic fixed in advance, the $p$-value $\Pr_{Z'}\{V(Z',Y_B)\ge V(Z,Y_B)\}$ over the design is valid under the null that $B$'s traffic does not depend on whether $G$ is throttled---whatever the benign traffic's law, drift or coordination.
\end{proposition}

This is Fisher's randomisation test \citep{fisher1935design,imbens2015causal} in a switchback design \citep{bojinov2023design}. In simulation on WildChat populations its null held on real, drifting traffic, and campaigns that reassigned throttled work to unflagged colluders were caught in every run, while those that did not lost 40\% of their throughput. But a benign team that shares a work queue and fails over legitimately fired the test 82--98\% of the time, throttling harms the legitimate accounts in a flagged group, and twelve weekly blocks take a quarter of a year. Whether such a test is specific in practice depends on how often benign groups respond to enforcement on other accounts---a rate only providers can measure. Appendix~\ref{app:randomized} gives details.

\section{Randomised interventions}
\label{app:randomized}

In pooled WildChat populations of 250, 2{,}500 and 10{,}000 account-weeks, each with its own recorded traffic in the 12 following weeks, a 50-account MetaMathQA campaign was injected, and the group chosen by multi-outlier selection was throttled in each of 12 weekly blocks independently with probability one half (conditioned on at least one throttled and one unthrottled block; 4{,}094 admissible schedules). The response was the on-task volume of the 50 accounts most correlated with the flagged group in the hollowed matrix, and the statistic, fixed before randomising, was the larger of the throttled-minus-unthrottled differences at lags 0 and 1. The null held on real, drifting traffic (one rejection at 1\% in 120 null runs). Where the flagged group held about half the colluders, a campaign that reassigned throttled work within the block or one block later was detected in every run; one that dropped the work lost 40\% of its throughput. At 10{,}000 accounts the flagged group already held all 50 colluders, leaving no one to fail over to. A benign team of 20 accounts with failover, half of it flagged, fired the test in 82--98\% of runs. At the deployed precision measured at 5{,}000--7{,}552 account-weeks (0.73 and 0.54), about 16--42 legitimate accounts would be throttled per detection. In a toy simulation of randomised answer marks, in which each answer carried one of four equivalent follow-up suggestions chosen at random, a campaign that drew later queries from suggestions shown to its other accounts was detected within four weeks 0.62 of the time at 10\% uptake and a fixed-prompt campaign never; a benign team that shares answers was detected exactly as often as the campaign, so the confound is structural.

\section{Guidance for providers in full}
\label{sec:guidance-full}

Table~\ref{tab:regime-full} summarises what worked where, and the list below turns the results into practice. Every item is backed by a measurement in this paper, and every item carries its caveat.

\begin{table}[t]
\caption{What works where: weekly detection of each campaign type by population and detector. Deployed thresholds in weekly populations; oracle thresholds in pooled ones, where no deployed threshold reached 1\% false alarms.}
\label{tab:regime-full}
\begin{center}
\small
\begin{tabular}{llccc}
\toprule
Population & Campaign & Two-scale & Hollowed & Multi-outlier \\
\midrule
Development weeks, 64--364 accounts & Dataset & 0.79--0.85 & 1.00 & 1.00 \\
\quad (deployed) & Half padding & 0.21 & 0.44--0.48 & 0.36--0.42 \\
 & Least detectable & 0.00 & 0.01 & 0.01 \\
Untouched weeks, 64--252 accounts & Dataset & 0.86 & 1.00 & 0.97 \\
\quad (deployed) & Half padding & 0.00 & 0.50--0.57 & 0.00 \\
 & Least detectable & 0.00 & 0.04 & 0.00 \\
Pooled, 1{,}000 accounts, 5\% colluders & Dataset & 0.75--0.80 & 1.00 & 1.00 \\
\quad (oracle) & Half padding & 0.00 & 0.07 & 1.00 \\
Pooled, 10{,}000 accounts, 0.5\% colluders & Dataset & 0.00 & 0.03 & 1.00 \\
\quad (oracle) & Half padding & 0.00 & 0.03 & 0.17 \\
\bottomrule
\end{tabular}
\end{center}
\end{table}

\begin{enumerate}[leftmargin=1.5em]
\item \textbf{Measure your benign coordination first.} Compute, per week, the real-to-shuffled ratio of the top eigenvalue of the centred hollowed matrix, the number of outlier eigenvectors, and the share of accounts of each valuable type. These numbers decide what any query-only monitor can see, and the eigenvalue ratio changed by a factor of two between two periods of the same public log. A short script that computes them from embeddings, account identifiers and week labels is in the code release (\coderepo).
\item \textbf{Calibrate on whole historical windows against a lagged reference.} Never build the null by shuffling accounts or queries. Expect realised false alarms of about twice the nominal rate.
\item \textbf{Protect the reference and the calibration history.} A campaign that ramps up through the calibration weeks poisons both, and it sharply cut the detection of every detector we built. Keep a frozen reference and a frozen calibration set beside the rolling ones, and audit how thresholds drift.
\item \textbf{Monitor whole populations, and select on the hollowed matrix.} Sharding defeats every detector. Hollowed selection was the most robust detector on untouched weeks; multi-outlier selection helps at scale but pays for multiplicity, so run both.
\item \textbf{Accumulate evidence over weeks} with a conformal test martingale; mimicry slows detection rather than preventing it, as long as the gap to benign traffic is positive.
\item \textbf{Combine with linkage in both directions.} A firing test is a high-precision pointer, not an enumeration of the campaign. From a partial metadata linkage, expanding along the hollowed matrix recovered nearly all remaining colluders.
\item \textbf{Expect evaluation pipelines in the alarms.} The monitor detects coordinated topic shift, not extraction. Use it as a filter for human review.
\item \textbf{Consider randomised enforcement.} Randomising when flagged groups are throttled turns every enforcement action into an exactly calibrated test of failover; whether a detected failover points to a campaign depends on how often your benign customers fail over, which only you can measure.
\end{enumerate}

\section{Related work in full}
\label{sec:related-full}

\paragraph{Model extraction and its defences.} Extraction through prediction APIs \citep{tramer2016stealing}, knockoff networks \citep{orekondy2019knockoff} and active-learning extraction \citep{pal2020activethief} established the threat; distillation into small students from generated instruction data \citep{hinton2015distilling,taori2023alpaca,yu2024metamath,wei2024magicoder} is its large-language-model form. Most deployed defences analyse one account at a time: PRADA \citep{juuti2019prada}, SEAT \citep{zhang2021seat} and HODA \citep{sadeghzadeh2023hoda}; stateful detection \citep{chen2020stateful} and Blacklight \citep{li2022blacklight} keep query histories and discuss attackers who split work across accounts; \citet{kesarwani2018model} track feature-space coverage. FDINet tests pairs of clients for collusion \citep{yao2023fdinet}; DAWN notes that Sybil accounts can be linked through query patterns and IP addresses \citep{szyller2021dawn}. Closest to us, \citet{liu2026embarrassingly} apply a benign-calibrated MMD test to traffic windows of sentence embeddings, including windows that mix campaign queries with cover queries; it is our strongest baseline, and we extend that line to whole real populations, colluder shares down to 0.5\% and a measured price of evasion. Other defences avoid detection altogether: honeypots steer extractors toward low-value knowledge \citep{dai2026knowledgetrap} or plant backdoors that reveal a copy \citep{wang2025honeypotnet}, and LADS makes each generation's randomness derive from a private seed so that multi-account distillation degrades \citep{diao2026lads}.

\paragraph{Prices of evasion.} Measured costs of evading a detector are not new. \citet{liu2026embarrassingly} found that adaptive query selection reduced their detector's power while cutting the student's accuracy from 76.4\% to about 67\%; SEAT forces adaptive attackers to use 3.8--16 times more accounts \citep{zhang2021seat}; FDINet evaluates adaptive feature-correction attacks and colluding clients \citep{yao2023fdinet}. Our addition is the price in an extraction regime on real traffic, its dependence on whether the student already has the skill, and its time-integrated form. \citet{gudibande2024false} found that imitating proprietary models transfers style far more than capability, which matches our elicitation regime.

\paragraph{Coordination, mimicry and poisoning.} Detecting groups of accounts that act together is a long-standing problem in fraud and bot detection: EigenSpokes \citep{prakash2010eigenspokes}, CopyCatch \citep{beutel2013copycatch} and Fraudar \citep{hooi2016fraudar} find lockstep groups through singular vectors or dense blocks; DeBot \citep{chavoshi2016debot}, SynchroTrap \citep{cao2014synchrotrap} and point-process models \citep{sharma2021identifying} use timing; ban-evasion detection links replacement accounts to banned ones \citep{niverthi2022characterizing}. Benign coordination is a documented confounder for coordination detectors \citep{vargas2020detection}. Mimicry attacks on intrusion detectors \citep{wagner2002mimicry} and boiling-frog poisoning of PCA anomaly detectors \citep{rubinstein2009antidote} are the ancestors of our padding and ramp attacks. PCA traffic-anomaly detection \citep{lakhina2004diagnosing} separates normal and residual subspaces, a role our lagged whitening plays.

\paragraph{Statistics.} Our spectral analysis rests on the BBP transition and its finite-rank extensions \citep{baik2005phase,paul2007asymptotics,benaych2011eigenvalues} and Tracy--Widom fluctuations \citep{tracy1994level}. Higher Criticism \citep{donoho2004hc} is the standard test for sparse weak signals. Statistical--computational gaps in sparse PCA, submatrix and planted-community detection \citep{berthet2013optimal,ma2015computational,hajek2015computational} frame our selection gap. The capture of top eigenvectors by individually heavy accounts is the localisation of edge eigenvectors on high-degree vertices \citep{alt2023localized,hiesmayr2023spectral}, whose standard repairs are regularisation and non-backtracking operators \citep{le2017concentration,krzakala2013spectral,saade2014spectral}; hollowing belongs to this family. Proposition~\ref{prop:lfc} is the I-projection \citep{csiszar1975idivergence} read as an attack, and its capacity form is the square-root law of steganography and covert communication \citep{ker2007capacity,bash2013limits}. Our sequential tools are e-values and conformal test martingales \citep{vovk2021evalues,vovk2021testing,vovk2021retrain,ramdas2023gametheoretic}.

\section{Limitations in full}
\label{sec:limitations-full}

We have tried to state each limit where its result appears. Collected here, they fall into five groups.

\paragraph{Data.} WildChat is consumer chat through one provider, not API traffic, and hashed IPs only approximate accounts: shared addresses merge users and VPNs split them, which may itself manufacture some of the templated ``bulk users'' that act as benign confounders. Our activity thresholds exclude low-volume accounts, which is where a patient attacker would hide. Weekly populations hold a few hundred accounts, with colluders at 14--50\% of a window; production shares are orders of magnitude lower. Real API logs, which only providers hold, are the decisive next measurement.

\paragraph{Representation.} Every real-traffic result is detectability in one family of sentence embeddings: MiniLM, and E5-small-v2, which is initialised from MiniLM. Attacker and defender therefore share a representation. Lexical, perplexity, style, near-duplicate and timing features, and the victim model's own hidden states, are untested here. Our own discriminator separated LLM-written prompts from real ones with AUC 0.97--0.98, but we did not use it as a detector.

\paragraph{Statistics.} Week-level conclusions rest on 17--24 test weeks; run-to-run variation exceeds the intervals of single runs; low-false-alarm results rest on few null exceedances; and our development-week results were obtained after adaptive reuse of the same test weeks, which the single frozen evaluation only partly corrects. Predictions were locked in notebooks before each run but carry no external timestamp, so they are not pre-registrations in the formal sense.

\paragraph{Campaign construction.} Natural, tilted and written colluder accounts were assembled from draws across the attacker's pool rather than from whole real user-weeks, so they are less topical than real users; random real prompts assembled this way ranked above benign windows under hollowed selection (AUC 0.88). This confounds the natural and covert results in an unknown direction. We did not simulate relays.

\paragraph{Capability.} The frontier rests on one 360M student, one benchmark, answers taken from datasets and logs rather than a live teacher, and two seeds; at the covert end, differences are 4--18 problems out of 400. The writer that tried to imitate users was a 1.5B model; a frontier model might write far better. Larger students fell into the elicitation regime, where stealth is free.

\paragraph{Open tests.} The experiments that would most change our conclusions are: an anytime-valid sequential test on the hollowed score against padded and covert campaigns; covert and natural colluders built from whole real user-weeks; a relay simulation with and without session interleaving and selection; a defence against slow calibration poisoning; distillation with a live teacher, more seeds, the full test set and a second student and task; and, above all, the benign-coordination census on a real provider's logs.

\clearpage
\section{Implementation details}
\label{app:impl}

\begin{table}[ht]
\centering
\caption{Main settings.}
\small
\begin{tabular}{p{0.32\textwidth}p{0.6\textwidth}}
\toprule
Embedding & CIFAR-10: victim penultimate layer, PCA-32. WildChat: all-MiniLM-L6-v2, PCA to $d\in\{32,64,128,384\}$, rescaled to total variance 4. \\
Fourier modes & $m=400$ complex modes (CIFAR-10 and the small-population WildChat protocol); $m=300$ modes with augmented real whitening (realistic protocol). Bandwidths $h\in\{2,4,8\}$ with Bonferroni correction (CIFAR-10), $h=4$ (WildChat), and $h=4$ coarse with $h=0.5$ fine for the two-scale test. \\
Whitening floor & eigenvalues below $10^{-4}$ of the maximum discarded. \\
Reference & CIFAR-10: 300{,}000 benign queries. WildChat: up to 100{,}000 benign turns from the four preceding weeks (lagged). \\
Calibration & CIFAR-10: 1{,}000 null worlds per configuration, Bonferroni over bandwidths. WildChat: 40 null windows per week, thresholds from the 8 preceding weeks (realistic protocol) or 250 windows from the 4 preceding weeks (small populations); conformal thresholds over all prior weeks for hollowed and multi-outlier selection. \\
MMD baseline & unbiased MMD$^2$ on normalised MiniLM embeddings, Gaussian kernel with median-heuristic bandwidth, window sample of 2{,}000 queries against 20 reference batches of 500. \\
Fraudar baseline & account $\times$ topic count matrix (100 $k$-means topics), column weights $1/\log(\text{degree}+5)$, greedy peeling. \\
Distillation & Qwen2.5-0.5B (base), LoRA $r=16$, $\alpha=32$, dropout 0.05 on all attention and MLP projections; one epoch, batch 8, learning rate $2\times10^{-4}$, maximum length 512; GSM8K test problems 1--500, greedy decoding, last-number match. \\
Elicitation students & Qwen2.5-0.5B as above on 100--900 queries, three seeds, GSM8K problems 1--400. Qwen2.5-1.5B (base) on MATH-500: fp16 base with fp32 LoRA ($r=16$), micro-batch 2 with 4 accumulation steps, learning rate $2\times10^{-4}$, maximum length 512, up to 512 new tokens, boxed-answer match after normalisation. \\
Extraction student & SmolLM2-360M (base), full fine-tuning in fp16 autocast, AdamW, learning rate $5\times10^{-5}$, no weight decay, linear schedule with warm-up over the first 1/30 of steps, effective batch 16 (micro-batch 4), one pass, maximum length 512; prompt format ``Question: \ldots\ Answer:'', loss on answer tokens only; GSM8K problems 1--400, greedy decoding, up to 320 new tokens, last-number match; two seeds. \\
Rewriter & Qwen2.5-1.5B-Instruct, sampling at temperature 0.8 (0.9 in regeneration rounds), top-$p$ 0.95, up to 110 new tokens; shown three real on-task user messages as style examples; 5{,}000 MetaMathQA GSM-type problems; two regeneration rounds for rewrites the discriminator caught. \\
Classifiers & Logistic regression on MiniLM embeddings, balanced class weights. Task region: MetaMathQA (or human GSM8K training questions, for the value-defined region) against four times as many benign WildChat turns from the first 8 weeks; region = posterior $\ge0.5$. Discriminator: two-fold cross-fitted, rewrites against real on-task prompts; a rewrite passes if its held-out posterior is below 0.5. \\
Compute & Kaggle notebooks with one T4 or P100 GPU; each experiment runs in 0.5--2 hours. \\
\bottomrule
\end{tabular}
\end{table}

\FloatBarrier
\section{Reproducibility}
\label{app:repro}

Every number comes from Kaggle notebooks run with fixed seeds on one T4 or P100 GPU. Experiments ran as numbered steps, each in its own notebook, and before each run we wrote its predictions into the notebook header; they carry no external timestamp, so we call them locked rather than pre-registered. Of 184 predictions, 75 passed, 30 passed partially, 74 failed and 5 could not be tested. The code at \coderepo{} contains every notebook; a map from each table, figure and quoted number in this paper to the script and line that computes it; the full prediction log with outcomes; the corrections we made during the study; the figure script; and the benign-coordination census script. CIFAR-10, WildChat-1M, WildChat-4.8M, MetaMathQA, Magicoder-OSS-Instruct, Alpaca, GSM8K, MATH-500 and MMLU are public; the WildChat notebooks require a Hugging Face token with access to the datasets.

\end{document}